%% file: paper.tex
\documentclass[conference, letterpaper]{IEEEtran}
\usepackage{adjustbox}
\usepackage{enumerate}
\usepackage{graphicx}
\usepackage{epstopdf}
\usepackage{epsfig}
\usepackage{color}
\usepackage{url}
\usepackage{etoolbox}
\usepackage{booktabs}
\usepackage{amsmath}
\usepackage{amssymb}
\usepackage{mathtools}
\usepackage{hhline}
\usepackage{setspace}
\usepackage{balance}
\usepackage{xcolor,colortbl}
\usepackage{array}
\usepackage[ruled, noend]{algorithm2e}
\usepackage{blindtext}
\usepackage{multirow}
\usepackage{mathrsfs}
\usepackage{tabularx}
\usepackage{cleveref}
\usepackage{spverbatim}
\usepackage{float}
\usepackage{caption}
\usepackage{tabularx}
\usepackage{cite}
\usepackage{subcaption}
\usepackage{cleveref}
\usepackage[utf8]{inputenc}
\usepackage{todonotes}

\crefname{section}{\S}{\S}

\definecolor{LightCyan}{rgb}{0.88,1,1}
\definecolor{Gray}{gray}{0.9}
\definecolor{ashgrey}{rgb}{0.7, 0.75, 0.71}

\newcommand{\descr}[1]{\smallskip\noindent\textbf{#1}}
 
\IEEEoverridecommandlockouts

\begin{document}

\title{A Streaming Machine Learning Framework for Online Aggression Detection on Twitter}

\author{
	\IEEEauthorblockN{Herodotos Herodotou}
	\IEEEauthorblockA{
		\textit{Cyprus University of Technology}\\
		Limassol, Cyprus\\
		herodotos.herodotou@cut.ac.cy}
	\and
	\IEEEauthorblockN{Despoina Chatzakou}
	\IEEEauthorblockA{
		\textit{Centre for Research and Technology Hellas}\\
		Thessaloniki, Greece\\
		dchatzakou@iti.gr}
	\and
	\IEEEauthorblockN{Nicolas Kourtellis}
	\IEEEauthorblockA{
		\textit{Telefonica Research}\\
		Barcelona, Spain\\
		nicolas.kourtellis@telefonica.com}
}

\maketitle

\input{abstract}


\input{sections/introduction}

\input{sections/related_work}
\input{sections/system}

\input{sections/dataset}

\input{sections/experiments}
\input{sections/conclusion}

\section*{Acknowledgments}
The research leading to these results has received funding from the EU's H2020 Programme, No 691025 (MSCA-RISE project ENCASE) and No 830927 (project CONCORDIA).
The paper reflects only the authors’ views and the Commission is not responsible for any use that may be made of the information it contains.

\balance

\bibliographystyle{IEEEtran}
\bibliography {paper}

\end{document}

%% file: abstract.tex
\begin{abstract}

The rise of online aggression on social media is evolving into a major point of concern.
Several machine and deep learning approaches have been proposed recently for detecting various types of aggressive behavior.
However, social media are fast paced, generating an increasing amount of content, while aggressive behavior evolves over time.
In this work, we introduce the first, practical, real-time framework for detecting aggression on Twitter via embracing the streaming machine learning paradigm.
Our method adapts its ML classifiers in an incremental fashion as it receives new annotated examples and is able to achieve the same (or even higher) performance as batch-based ML models, with over 90\% accuracy, precision, and recall.
At the same time, our experimental analysis on real Twitter data reveals how our framework can easily scale to accommodate the entire Twitter Firehose (of 778 million tweets per day) with only 3 commodity machines.
Finally, we show that our framework is general enough to detect other related behaviors such as sarcasm, racism, and sexism in real time.

\end{abstract}

%% file: sections/introduction.tex

\section{Introduction}
\label{sec:introduction}

In the last years, online aggression has spiked across the Web, with many incidents of abusive behavior being reported in different contexts~\cite{cyberbullying-facts,passive-aggression}.
Aggressive behavior can be observed in any type of platform, independently of the audience it aims to engage with respect to age, gender, etc., and the utility or service the specific platform offers.
In particular, aggressive, abusive, offensive, bullying, racist, or sexist behavior has been studied in different online contexts: cyberbullying and aggression on Twitter and other media (e.g.,~\cite{dinakar2011modeling, meanBirds,chatzakou2019detecting, founta2018large, riadi2017detection}), hateful speech (e.g.,~\cite{waseem2016hateful, davidson2017automated}), and offensive or abusive language (e.g.,~\cite{lee2002holistic, Chen2012DetectingOffensiveLanguage, nobata2016abusive}), to name a few.
This problem has been studied in different social media platforms such as Twitter~\cite{meanBirds,chatzakou2019detecting}, Instagram~\cite{Hosseinmardi2015,yao2019cyberbullying}, YouTube~\cite{Chen2012DetectingOffensiveLanguage}, Yahoo Finance~\cite{Djuric2015HateSpeechDetection}, Yahoo Answers~\cite{kayes2015ya-abuse}, 4chan~\cite{4chan}, and across various demographics (e.g., teenagers vs. adults, men vs. women, etc.~\cite{Smith2008CyberbullyingNature,cyberbullying-facts}).
Overall, it is difficult to find a general definition of aggression, as this term is often used interchangeably to describe different types of abnormal behavior.
To this end, here, we define as aggressive behavior the existence of abusive language and hate speech in online content; could be an one-off or repetitive behavior that could, for instance, lead to bullying.

With all the negative publicity this problem regularly receives in the news, popular online platforms have tried to curb the problem by deploying new features and methods.
For example, Twitter~\cite{bbc2020Twitter-bullying-feature} and Instagram~\cite{instagram2019Bullying-feature} have recently proposed new features that aim to limit or block the ability of aggressors to bully by posting messages on their victims' pages, profiles, or walls.
However, even with all this effort in place, the problem still persists.
This can be attributed to the following aspects of the problem.

First, many of these online platforms enable users to engage in a fast-paced generation of content across different topics, thus, facilitating an exponential growth of user-contributed posts:
$2.5$ trillion posts to date on Facebook~\cite{social-media-stats}; 
Instagram users hitting the ``like'' button $4.2$ billion times per day~\cite{social-media-stats}; 
more than $2$ million ``snaps'' created every minute on Snapchat~\cite{social-media-stats};
$778$ million tweets posted per day on Twitter~\cite{internet-live-stats}; etc.
Therefore, the platforms have a constantly increasing workload to process for detecting abusive behavior.
Second, users (and especially millennials, Generation Z, and even Generation Alpha) adapt quickly to new community rules or algorithms of platforms rolled out to detect and ban such behavior.
In fact, they typically find ``innovative'' ways to circumvent the rules in order to achieve their aggressive goals, by using new words or special text characters to signify their aggression but avoid detection~\cite{social-norms-chi18}.
Third, this complex behavior is difficult to model, as it highly depends on the context, cultural background of users, nature of user interaction, platform features, scope and utility, etc.
Finally, detecting this behavior is a minority class problem, making it harder to address it effectively with automated systems, as there are not many examples to train classifiers on, at any given random sample of data.

Unfortunately, past academic and industry efforts to study the problem and propose machine learning (ML) algorithms to detect such behavior have not addressed the above aspects in an effective manner.
As most current ML approaches used are batch-based, a trained model can become deprecated after some time~\cite{rafiq2018scalable}, while many approaches are computationally expensive for both training and testing purposes~\cite{nobata2016abusive, zhang2018detecting}.
It becomes apparent that there is urgent need for new, real-time ML methods that:
1) \textit{adapt} their ML modeling in a similarly fast rate as the content becomes available, keeping the ML models up-to-date with users' adaptability, and
2) can \textit{scale} to process the increasing workload from generated content in real time.

In this work, we address this gap, and introduce the \textit{first, practical, and real-time framework for aggression detection on Twitter}.
Departing from past works, and embracing the streaming ML modeling paradigm, our method adapts its ML classifiers in an incremental fashion as it receives new annotated examples.
Moreover, we argue on the framework's ability to easily scale out and accommodate the whole Twitter Firehose stream with a handful of machines.

In summary, our \textbf{contributions} are as follows:
\begin{itemize}
\item We propose the first \textit{real-time detection framework} for aggression on Twitter.
We designed it to be extensible, scalable, and accurate, thus being able to provide useful alerts to human moderators.
We employ state-of-the-art streaming ML methods such as Hoeffding Trees, Adaptive Random Forests, and Streaming Logistic Regression.

\item We propose an \textit{adaptive bag-of-words method} for improving the ML modeling performance while new aggressive cases appear.
This feature allows the framework to further adapt in transient aggressive behaviors such as the use of new (key)words for spreading abuse or hate speech.

\item We design our framework to be \textit{deployable on state-of-art streaming engines} such as Apache Spark and discuss how it can be extended and run on other streaming platforms.

\item We \textit{demonstrate its effectiveness} to detect aggressive behavior at real time on a large dataset of $86k$ tweets, annotated for this context.
We show that the results acquired from the streaming methods can be as good or even better than typical batch-based ML methods, with over $90\%$ performance in multiple metrics (accuracy, precision, recall).
We also perform an in-depth investigation of how factors such as preprocessing and normalization can affect streaming ML performance.

\item We test the pipeline on two other relevant, but smaller Twitter datasets, and demonstrate the method's ability to \textit{detect similar behaviors} with minimal adaptation and tuning.

\item We show how \textit{the framework can easily scale} to increase its throughput to over $14k$ tweets/sec and accommodate all the Twitter Firehose with only $3$ commodity machines.

\end{itemize}

\descr{Outline.}
The rest of this paper is organized as follows. 
Section \ref{sec:rel_work} reviews prior related work.
Section \ref{sec:system} presents the design and implementation of our framework, while
Section \ref{sec:datasets} discusses the dataset and the features employed for detecting aggression.
Section \ref{sec:experiments} describes our evaluation methodology and results on a real-world Twitter dataset.
Finally, Section \ref{sec:conclusion} discusses possible future directions and concludes the paper.

%% file: sections/related_work.tex

\section{Related Work}
\label{sec:rel_work}

Numerous studies have focused on the detection of abusive behavior on online social networks.
Under the umbrella of ``abuse'', various abnormal behaviors may be included, such as offensive language and hate speech, bullying and aggression, sexual exploitation, etc.
For instance, authors in~\cite{Chen2012DetectingOffensiveLanguage} proposed a method to detect highly probable users in producing offensive content in YouTube comments, while authors in~\cite{badjatiya2017deep} focused on hate speech detection, and specifically to detect racism and sexism.
More recently, a method for detecting bullying and aggressive behavior on Twitter has been proposed, where a wide range of profile, content, and network characteristics was considered for characterizing users' online behaviors~\cite{chatzakou2019detecting}.

Towards an efficient detection of abusive behavior, ML-based approaches are among the most commonly considered ones nowadays.
Many of the previous works used traditional ML classifiers, such as logistic regression~\cite{Djuric2015HateSpeechDetection, nobata2016abusive, waseem2016hateful, davidson2017automated}, support vector machines~\cite{warner2012detecting}, or ensemble classifiers~\cite{burnap2015cyber}.
For instance, authors in~\cite{nobata2016abusive} used an ML-based method to perform hate speech detection on Yahoo Finance and News data, while authors in~\cite{burnap2015cyber} used a combination of probabilistic, rule-, and spatial-based classifiers with a vote ensemble meta-classifier to study the spread of online hate speech on Twitter.

More recently, deep learning architectures have also been employed, in an effort to improve the detection efficiency.
For instance, in~\cite{badjatiya2017deep} various deep learning architectures were evaluated, including Convolutional Neural Networks (CNNs), Long Short-Term Memory Networks (LSTMs), and FastText~\cite{joulin2016bag}.
In the same direction, to detect hate speech on Twitter, authors in~\cite{park2017one} used deep learning models (CNNs), while in~\cite{zhang2018detecting} a CNN with Gated Recurrent Unit (GRU) network was used, combined with word embeddings.
Finally, for the detection of abusive and aggressive behaviors, authors in~\cite{chatzakou2019detecting} experimented with both traditional machine learning methods (i.e., probabilistic, tree-based, and ensemble classifiers), as well as deep neural networks, i.e., Recurrent Neural Networks (RNN), with the first ones resulting to a better performance.

\begin{figure*}[t]
	\centering
	\includegraphics[width=.75\textwidth]{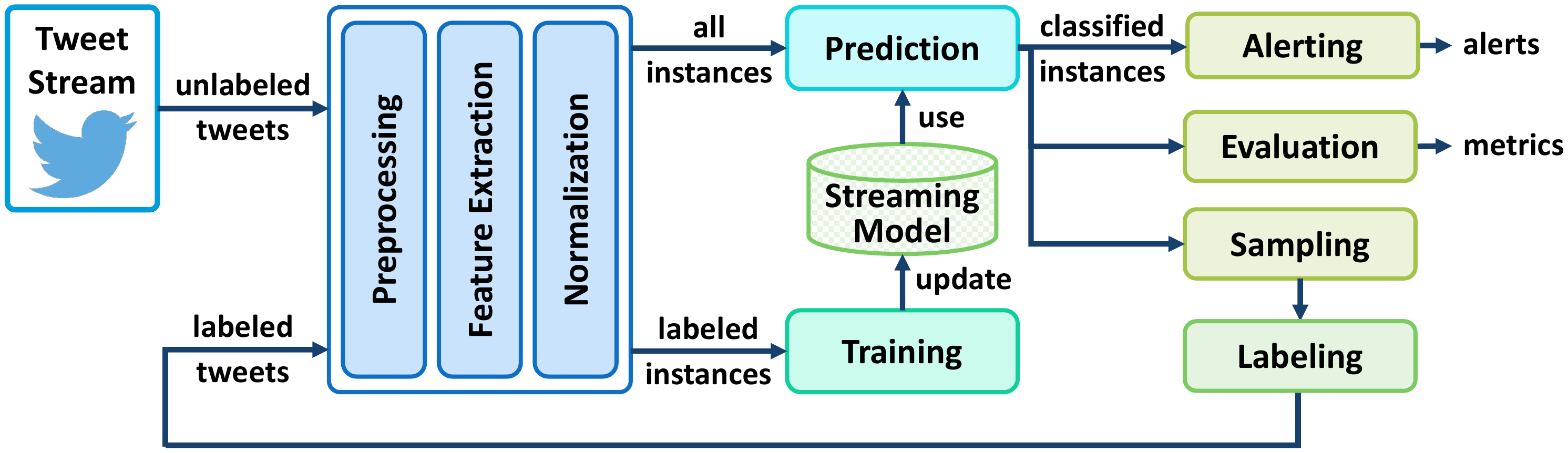}
	\caption{Overall system architecture for aggression detection on Twitter.}
	\label{fig:overall-workflow}
	\vspace{-4pt}
\end{figure*}

So far, most of the previous work has focused on the detection of abusive behaviors on a ``batch mode''.
That is, the objective was: given a set of data, to train an ML classifier to identify with the maximum possible effectiveness different types of abusive behaviors that are evident in the data.
To increase the accuracy of the detection process, some of the proposed approaches are cost-ineffective in terms of the time required for both training and testing purposes.
But as we live in an ever-evolving world, it is of paramount importance to develop methods that can support the continuous monitoring of the content produced in online sources, so that the detection of abusive behaviors can be performd in its infancy.
Towards this direction, authors in~\cite{rafiq2018scalable} proposed a multi-stage cyberbullying detection solution for Vine comments in order to reduce both the classification time and the time to raise alerts.
Specifically, an ``incremental computation'' approach was followed at both the feature extraction and classification stages, where data from previous stages was used within the current stage in order to lower the computational complexity.
Focusing on Instagram, authors in~\cite{yao2019cyberbullying} proposed a two-stage online framework for cyberbullying detection, where comments were examined as they became available.
To ensure the scalability of the detection process, an online feature selection approach was proposed for reducing the number of features used for classification (instead of constantly using the same set of features).
Unlike our approach that detects aggression on individual tweets, the two aforementioned approaches focus on detecting cyberbullying at the level of media sessions (i.e., for a set of comments).

\descr{Contributions.} 
To the best of our knowledge, the present study is the first to apply the streaming ML modeling paradigm for detecting aggressive or other forms of inappropriate behaviors on social media.
In fact, we extend our preliminary binary classifier~\cite{herodotou2021icde-streaming-aggression}, add a new adaptive feature, and attempt multi-class classification, as well as test our framework \emph{as is} on further datasets and classes.
Unlike existing works, our framework can adapt the ML models quickly and efficiently as content becomes available, while at the same time it can scale and process an increasing workload in real time.

%% file: sections/system.tex

\section{System Architecture}
\label{sec:system}

The overall goal of this work is to produce a detection framework for Twitter that is capable of:
(i) detecting aggressive behavior in real time as soon as new tweets arrive;
(ii) dynamically updating the classification model in real time as soon as new labeled tweets arrive; and
(iii) scaling to accommodate for the ever-increasing volume of tweets without sacrificing the classification performance. 
The first two are achieved via adopting the streaming data paradigm (i.e., computing on data directly as it is produced), while the latter is achieved by ensuring that almost all computations can be parallelized and executed in a distributed setting.
In the remainder of this section, we present the overall system architecture, discuss how our approach can scale out to handle the entire Twitter Firehose, and introduce a few of the supported state-of-the-art streaming classification methods.

\subsection{System Overview}
\label{sec:system:overview}

Figure~\ref{fig:overall-workflow} shows the overall system architecture for detecting aggressive behavior on Twitter, which involves the following steps:
(1) preprocessing, (2) feature extraction, (3) normalization, (4) training, (5) prediction, (6) alerting, (7) evaluation, (8) sampling, and (9) labeling.
The input consists of two streams with unlabeled and labeled tweets, while multiple outputs are possible, including alerts of aggressive tweets, evaluation metrics, and sample tweets for downstream labeling.

\descr{Data Input.}
The Twitter Streaming API constitutes the first input to our system and provides access to a stream of tweets in JSON format.
Each JSON payload contains information about the tweet (e.g., the textual content, posted timestamp, whether it is a retweet or a reply) and the user who posted the tweet (e.g., name, account creation timestamp, number of friends and followers).
The second input is a stream of labeled tweets using the same JSON format as the original tweets, plus an attribute containing the class label (e.g., normal, abusive, hateful; see Section~\ref{sec:datasets:data}).
With the exception of the training step that only applies to labeled tweets, all other steps process both unlabeled and labeled tweets in the same manner.

\descr{Preprocessing.}
The first step is to clean the tweet text by removing numbers, punctuation marks, special symbols (e.g., \$, \%, *) and URLs, as well as condensing white space.
In addition, we remove tweet-specific content, including known abbreviations (e.g., RT), hashtags, and user mentions.
Such preprocessing and cleaning are commonly used to simplify and improve the following feature extraction step.

\descr{Feature extraction.}
We extract an array of user profile, text, and social network features as detailed in Section \ref{sec:datasets:fextraction}.

\descr{Normalization.}
This step involves scaling attribute data to fall within a predefined range, such as between 0 and 1. 
We have implemented three common forms of normalization:
(i) \textit{minmax}, which uses the min and max values of each feature to scale a given value to be in the range [0-1];
(ii) \textit{minmax without outliers}, which rescales the min and max of each feature by removing statistical outliers,
before applying minmax normalization;
(iii) \textit{z-score}, which normalizes a given value to belong in a distribution with zero mean and unit standard deviation based on the mean and variance of the feature.	
The statistics required for normalization (e.g., min, max, mean) can be provided as input or they can be easily computed incrementally during the data stream processing.
As the performance of some streaming classification models is not affected by normalization, this step is optional.

\descr{Training.}
Given a stream of labeled feature instances, the training step will incrementally update the underlying streaming classification model.
Unlike batch processing (where an instance is processed multiple times), in streaming ML each instance is only processed once, the model is updated, and then that instance is discarded.
Hence, it is not necessary to store a historical training set and periodically retrain the model as is typically done in a batch setting.
Another benefit of stream-based learning is the ability to deal with concept drift, which takes place when the data distribution evolves over time. Therefore, the streaming model remains always up-to-date.

\descr{Prediction.}
This step uses the streaming model to predict the label of each instance by computing the probabilities that a particular instance belongs to each class label.
Prediction is performed for both unlabeled and labeled instances for different reasons.
The former is used for detecting aggressive tweets and raising appropriate alerts.
The later is useful for evaluating the classification performance of the model by comparing the predicted labels with the true ones.

\descr{Alerting.}
Depending on the prediction's outcome, this step will raise an alert whenever an aggressive behavior is detected.
Given that these alerts are raised in real time, multiple options exist for how they can be handled:
(i) they can be sent to human moderators for evaluating the suspected tweet, or
(ii) an automatic warning can be posted on the tweet, or
(iii) the tweet can be automatically removed.
Given this methodology should be operational at the platform level (i.e., Twitter), it is also possible to maintain an alert history per user, which can be used for automatically suspending a user account when repeated offenses are detected.

\descr{Evaluation.}
The classification of labeled instances is used for computing important metrics, including the confusion matrix, accuracy, precision, recall, F1-score, and others.
Such metrics are used for evaluating the performance of the streaming model and revealing how good the predictions are.
The evaluation step also computes some interesting statistics regarding the unlabeled instances such as the distribution of predicted class labels and how they change over time.

\descr{Sampling.}
The key application for sampling is to select a (small) subset of tweets for manual inspection and labeling.
As aggressive tweets are typically a minority in real-world settings, random sampling would result in an imbalanced dataset.
One way to address this issue is to implement a boosted random sampling technique that utilizes the predicted label for boosting a random sample with tweets that are likely to be aggressive (without biasing the resulting dataset)~\cite{founta2018large}.

\descr{Labeling.}
The goal of this step is to generate new labeled tweets that can be used for improving the performance of the streaming model over time.
Labeling can be performed by specialized moderators or via a crowd-sourcing platform like CrowdFlower \cite{founta2018large}, and is beyond the scope of this paper.

\subsection{Scaling ML Streaming Methods}
\label{sec:system:scaling}

\begin{figure}[t]
	\centering
	\includegraphics[width=.49\textwidth]{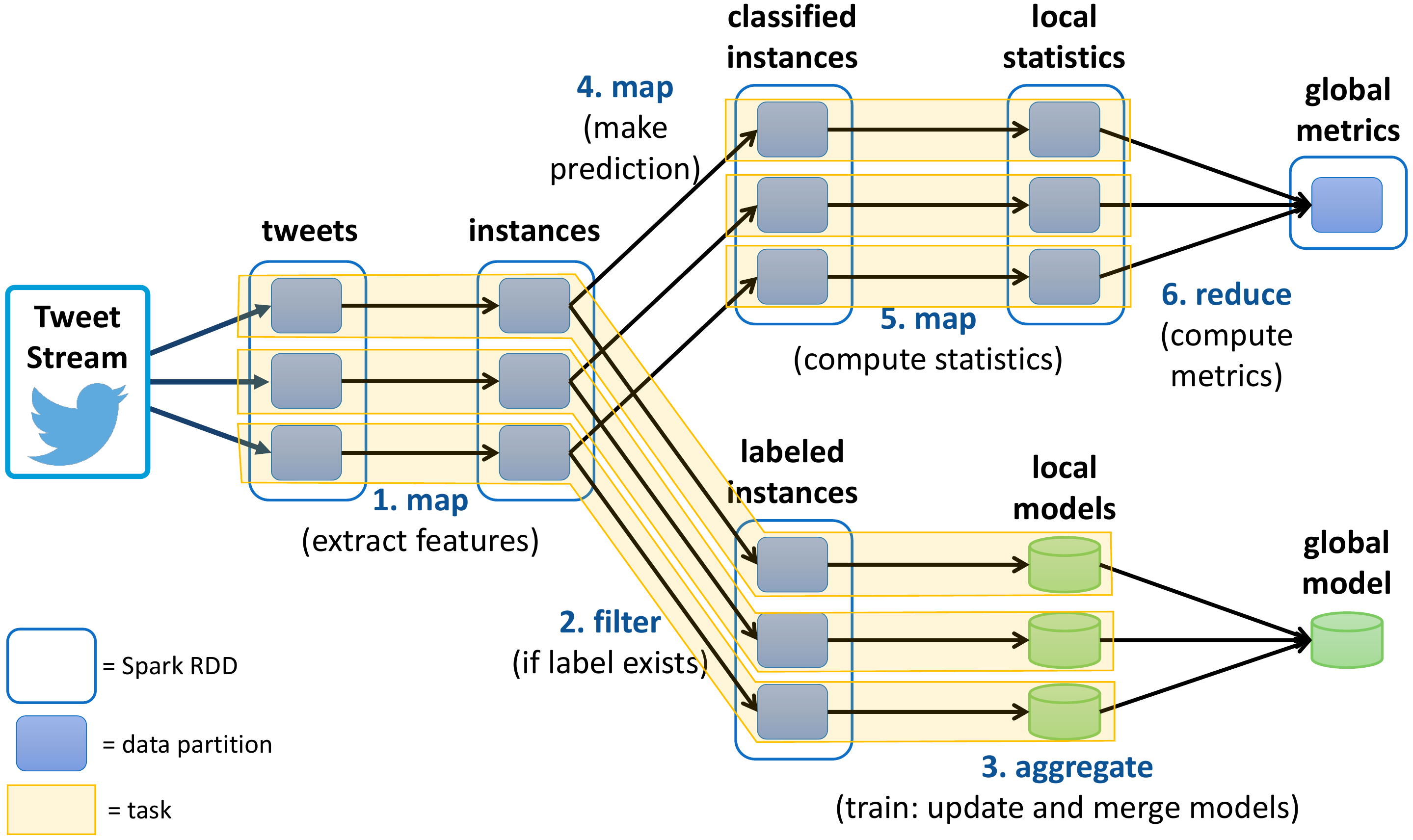}
	\caption{Dataset-oriented view of parallel task processing within a micro-batch in Apache Spark Streaming
	(alerting and sampling are executed in parallel starting from ``classified instances''; not shown for clarity).}
	\label{fig:microbatch-processing}
\end{figure}

Social media generate an enormous amount of data, on the order of several hundred GBs per day.
According to the latest statistics, there are over $778$ million tweets shared per day, amounting to $3.1$ TB of data (in JSON format) per day~\cite{internet-live-stats}.
Hence, for any online aggression detection solution to be practical, it must be able to scale to handle such massive data volumes and rates.

Currently, there are several distributed stream processing engines (DSPEs) that can process data in real time on clusters of commodity hardware, such as Apache Spark Streaming, Storm, Heron, and Flink.
We have implemented our framework on Spark Streaming~\cite{spark-streaming}, an extension of the Apache Spark platform that enables data stream processing over multiple sources.
Spark is a programmable framework for massive distributed processing of datasets, called Resilient Distributed Datasets (RDDs)~\cite{spark-rdd-nsdi12}.
In Spark Streaming, the underlying data abstraction is a discretized stream (DStream), represented as a sequence of RDDs.
Internally, the input data stream is divided into micro-batches,
with each micro-batch processed through the entire application at once using high-level transformations and actions such as map, filter, and aggregate.
Transformations on RDDs are divided and executed in parallel across the Spark cluster, leading to fast and scalable parallel processing.

Figure~\ref{fig:microbatch-processing} shows the key RDDs and transformations that are executed in parallel within a micro-batch in Spark Streaming.
The input tweets for one micro-batch are randomly divided into multiple data partitions that form the first RDD (named ``tweets'' in Figure~\ref{fig:microbatch-processing}).
Each input tweet is then mapped (op \#1) into an instance by applying the preprocessing, feature extraction, and normalization steps described above.
The labeled instances are then filtered (op \#2) before the training can take place using an aggregate operation (op \#3).
The training is performed in two parts; in the first one, the local model of each task is incrementally updated in parallel, while in the second part the local models are merged together to update the global model.
The map, filter, and the first part of aggregate are grouped together and executed using a set of parallel tasks.
The updated global model (with a serialized size of $<1$ MB) is then distributed across the cluster and is available for use by the tasks in the next micro-batch.
In addition, a new set of parallel tasks perform the predictions (op \#4) on all instances, followed by the computation of local statistics (op \# 5), such as the number of true positives, false positives, etc.
Finally, the local statistics are used to compute (op \# 6) global evaluation metrics such as accuracy, precision, and recall.
The alerting and sampling steps consume the ``classified instances'' RDD and are also performed in parallel (not shown in the figure).
Overall, almost all processing steps are fully parallelizable and can be executed on a cluster in an efficient and scalable manner (evaluated in Section~\ref{sec:exp-scalability}).

\begin{figure}[t]
	\centering
	\includegraphics[width=.49\textwidth]{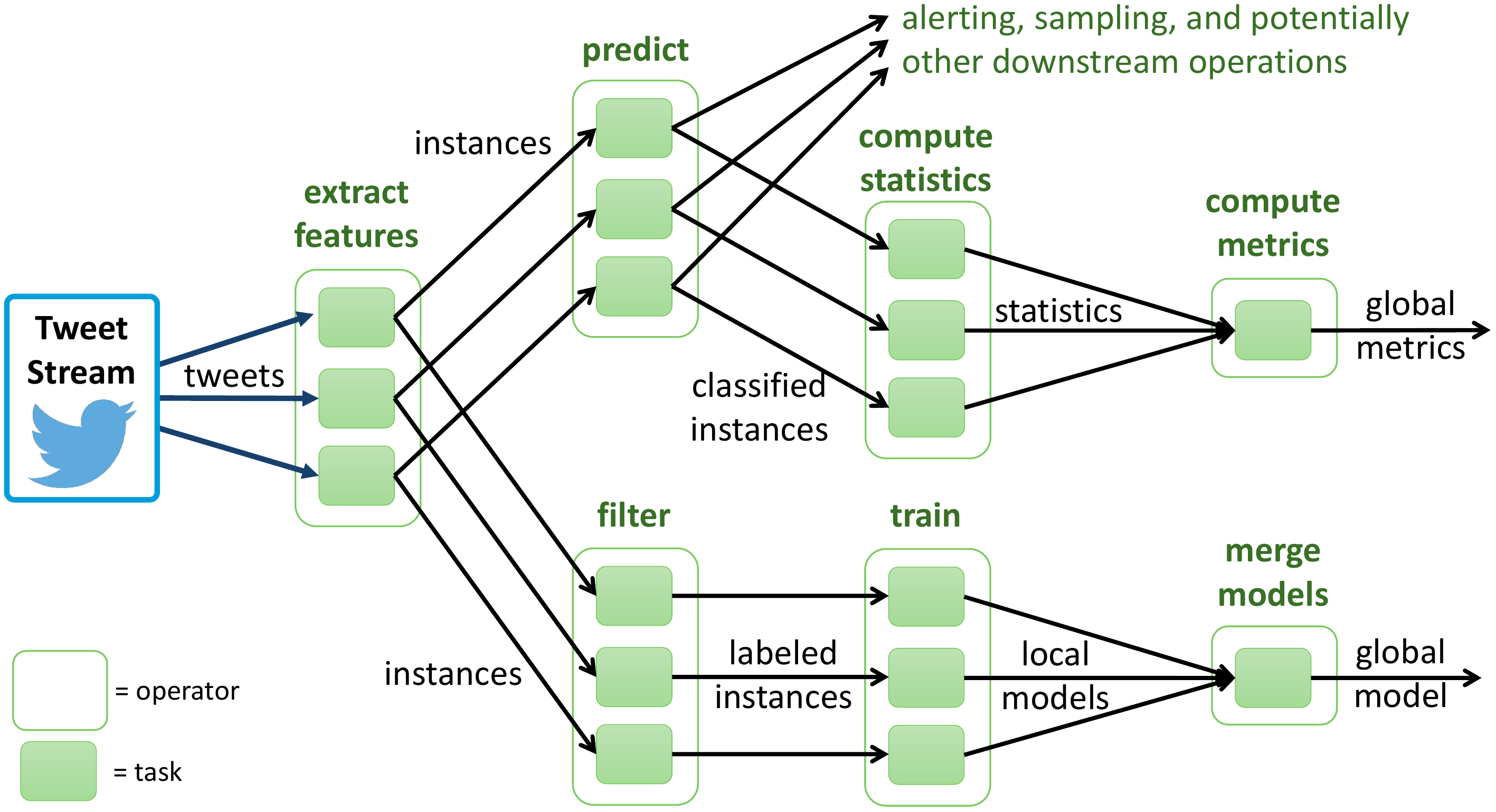}
	\caption{Task-oriented view of parallel processing in other distributed stream processing engines.}
	\label{fig:task-processing}
\end{figure}

Even though we have implemented our approach using Spark Streaming, the proposed architecture is general enough to be implemented in other DSPEs (e.g., Apache Storm, Heron, and Flink) that follow the per-record operator streaming model (as opposed to micro-batching).
These systems deploy a directed acyclic graph of operators that process the data stream in real time.
As shown in Figure~\ref{fig:task-processing}, an operator receives one or more streams, performs a streaming computation (e.g., map, filter, aggregation), and may output one or more streams.
Each operator can have multiple instantiations as individual tasks that run in parallel (Figure~\ref{fig:task-processing}).
Finally, each system has its own mechanism for sharing and periodically updating a shared state such as the global streaming model.

\subsection{Supported ML Streaming Methods}
\label{sec:system:methods}

Over the years, several online learning algorithms have been developed for performing classification, regression, clustering, and concept drift detection over evolving data streams.
Some frameworks such as Massive Online Analysis (MOA) \cite{moa-jmlr10} support streaming ML running on a single machine, while others such as Apache SAMOA \cite{samoa-website} implement several algorithms over various DSPEs (e.g., Apache Storm or Flink).
Similarly, \textit{streamDM} \cite{streamdm} is an open source software developed by Huawei for mining big data streams using Spark Streaming.
In our system, we have used three \textit{streamDM} classification algorithms that are designed for incremental training, namely Hoeffding Tree, Adaptive Random Forest, and Streaming Logistic Regression.

\descr{Hoeffding Tree (HT).}
The HT is an incremental decision tree learner designed for large data streams \cite{hoeffding-tree-kdd00}.
A tree node in a HT is expanded as soon as there is sufficient statistical evidence (based on the distribution-independent Hoeffding bound) that an optimal splitting feature exists.
The model learned by the HT is asymptotically nearly identical to the one built by a non-incremental learner, if the number of training instances is large enough.

\descr{Adaptive Random Forest (ARF) of HTs.}
ARF is an adaptation of the classical Random Forest algorithm that uses an effective resampling method based on online bagging and an updated adaptive strategy to cope with evolving data streams~\cite{arf-ml17}.
This ensemble classifier induces diversity through both online bagging and randomly selecting subsets of features for node splits.

\descr{Streaming Logistic Regression (SLR).}
Logistic Regression is a standard ML linear method that uses a logistic function to model a binary dependent variable (and can be easily extended to a multi-class scenario).
In the streaming setting, the fitting is similar to that performed offline, but with the model parameters updated online as new data arrives.
Finally, Stochastic Gradient Descent (SGD) is used to optimize the objective function.

%% file: sections/dataset.tex

\section{Datasets \& Feature Extraction}
\label{sec:datasets}

\subsection{Twitter Data}
\label{sec:datasets:data}
For our study, an abusive dataset provided by~\cite{founta2018large} is used, where tweets are characterized as \textit{abusive}, \textit{hateful}, \textit{spam}, or \textit{normal}.
The authors conducted several annotation rounds where different types of abusive behaviors were considered, i.e., offensive, abusive, hateful speech, aggressive, bullying, spam, and normal, to conclude to the aforementioned inappropriate speech categories.
Then, an exploratory study was carried out to determine the most representative labels related to the types of abusive content.
Overall, the dataset consists of $100k$ tweets.
Since our focus is on detecting aggressive behaviors in a streaming fashion, we removed the $14k$ tweets labeled as spam, as they can be handled with more specialized techniques such as~\cite{giatsoglou2015retweeting}.
Even though a dataset with $86k$ tweets is good for assessing the ML performance of the streaming algorithms, it is relatively small for testing the scalability of our approach.
Hence, using this dataset as seed, we also created a larger dataset with $2m$ unlabeled tweets for simulating the arrival of tweets in real time (see Section \ref{sec:exp-scalability}).

\subsection{Feature Extraction}
\label{sec:datasets:fextraction}

\begin{figure*}[!t]
	\centering
	\begin{subfigure}[b]{0.32\textwidth}
		\captionsetup{font=small}
		\includegraphics[width=\textwidth]{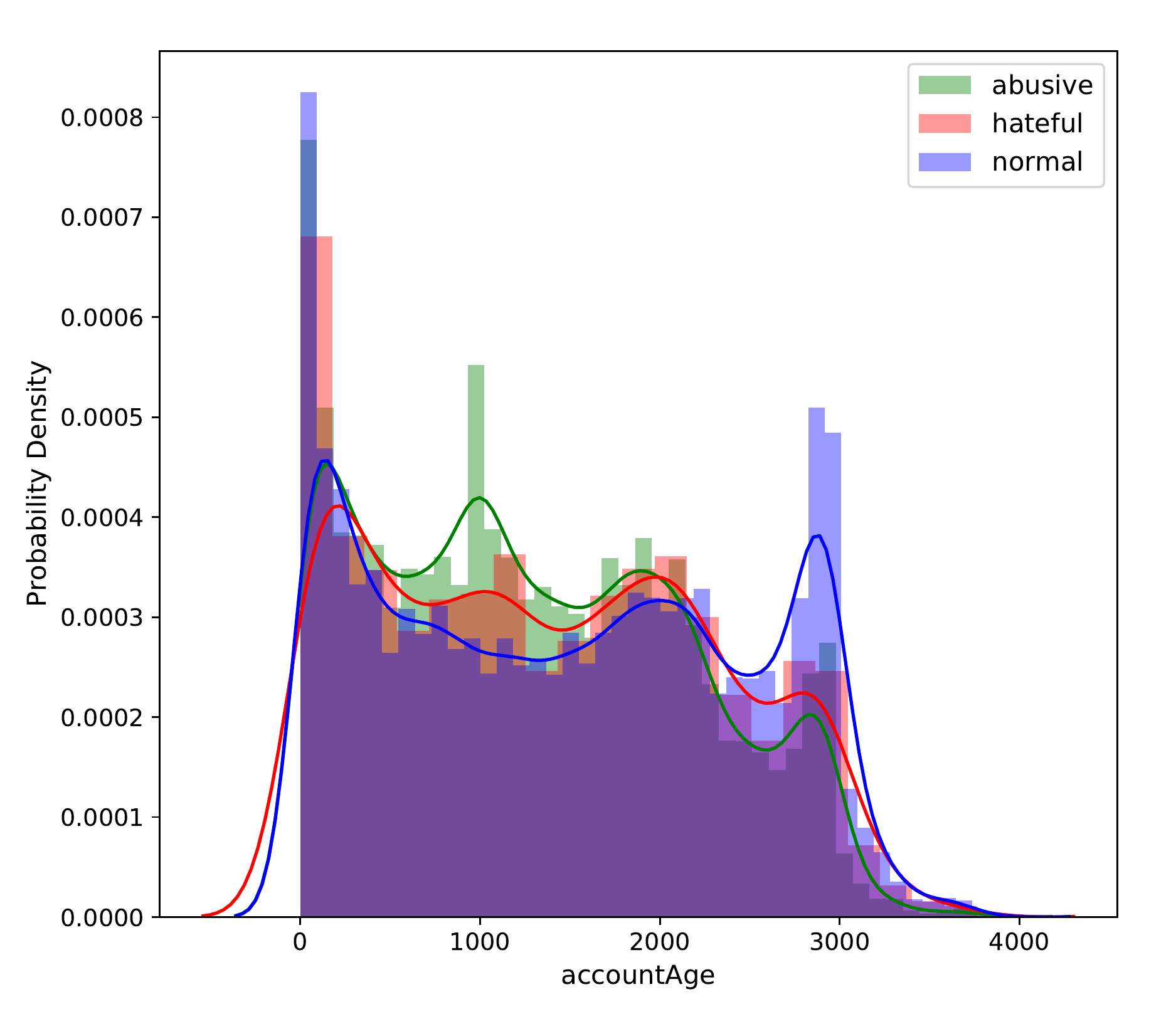}
		\caption{Account age (days).}
		\label{fig:accountAge}
	\end{subfigure}
	\begin{subfigure}[b]{0.32\textwidth}
		\captionsetup{font=small}
		\includegraphics[width=\textwidth]{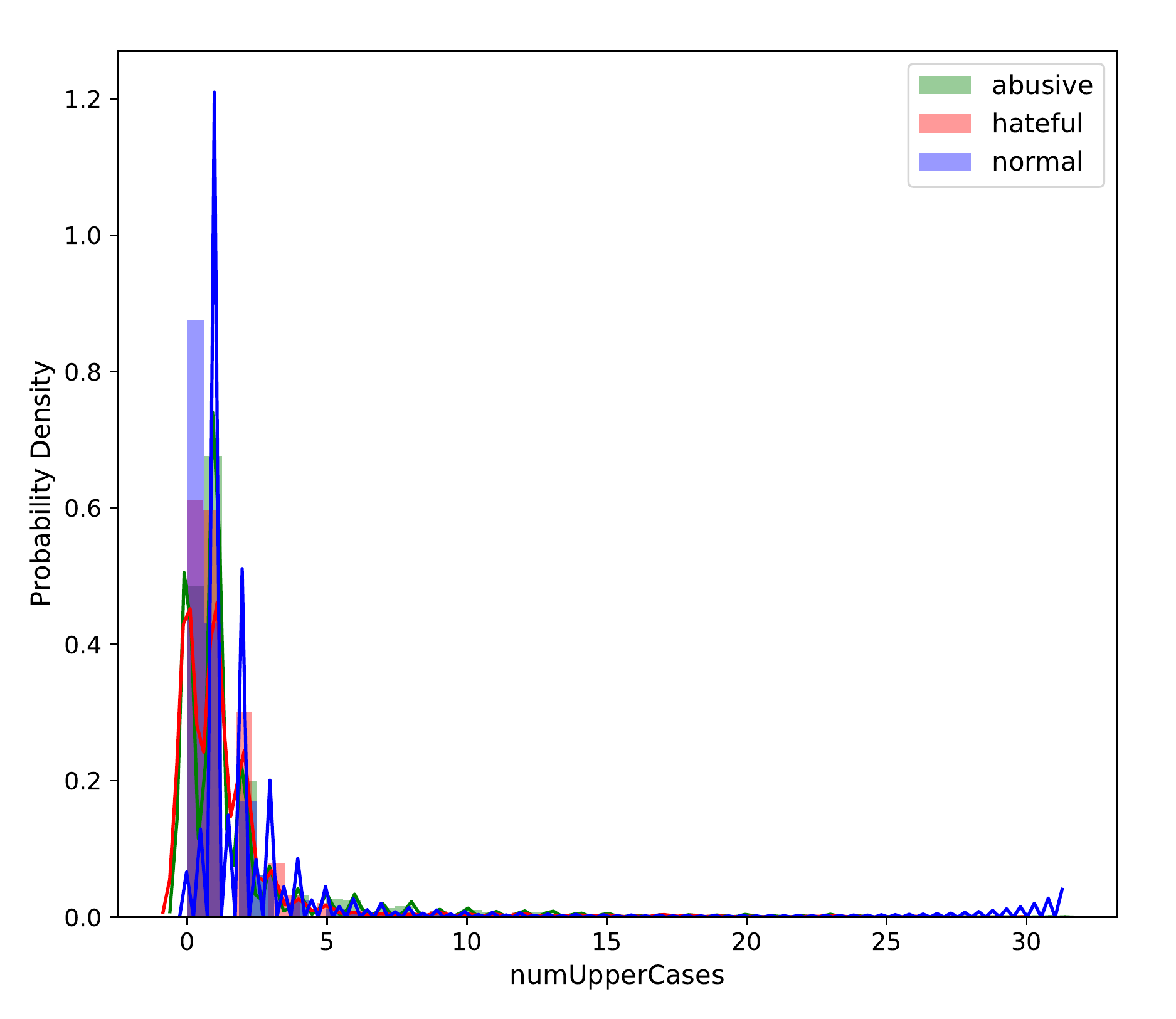}
		\caption{Uppercase letters.}
		\label{fig:uppercases}
	\end{subfigure}
	\begin{subfigure}[b]{0.32\textwidth}
		\captionsetup{font=small}
		\includegraphics[width=\textwidth]{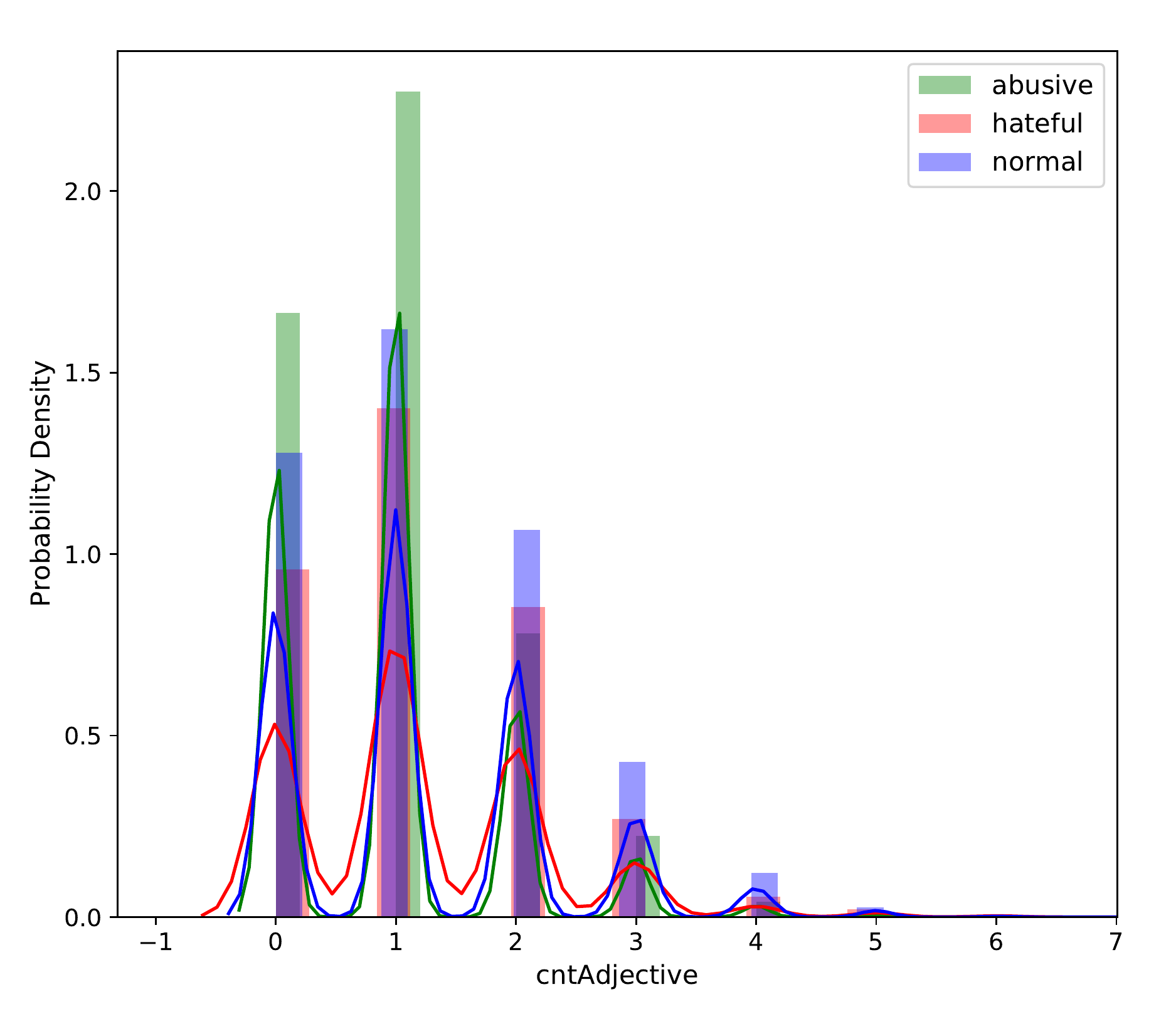}
		\caption{Adjectives.}
		\label{fig:adjectives}
	\end{subfigure}
	
	\begin{subfigure}[b]{0.32\textwidth}
		\captionsetup{font=small}
		\includegraphics[width=\textwidth]{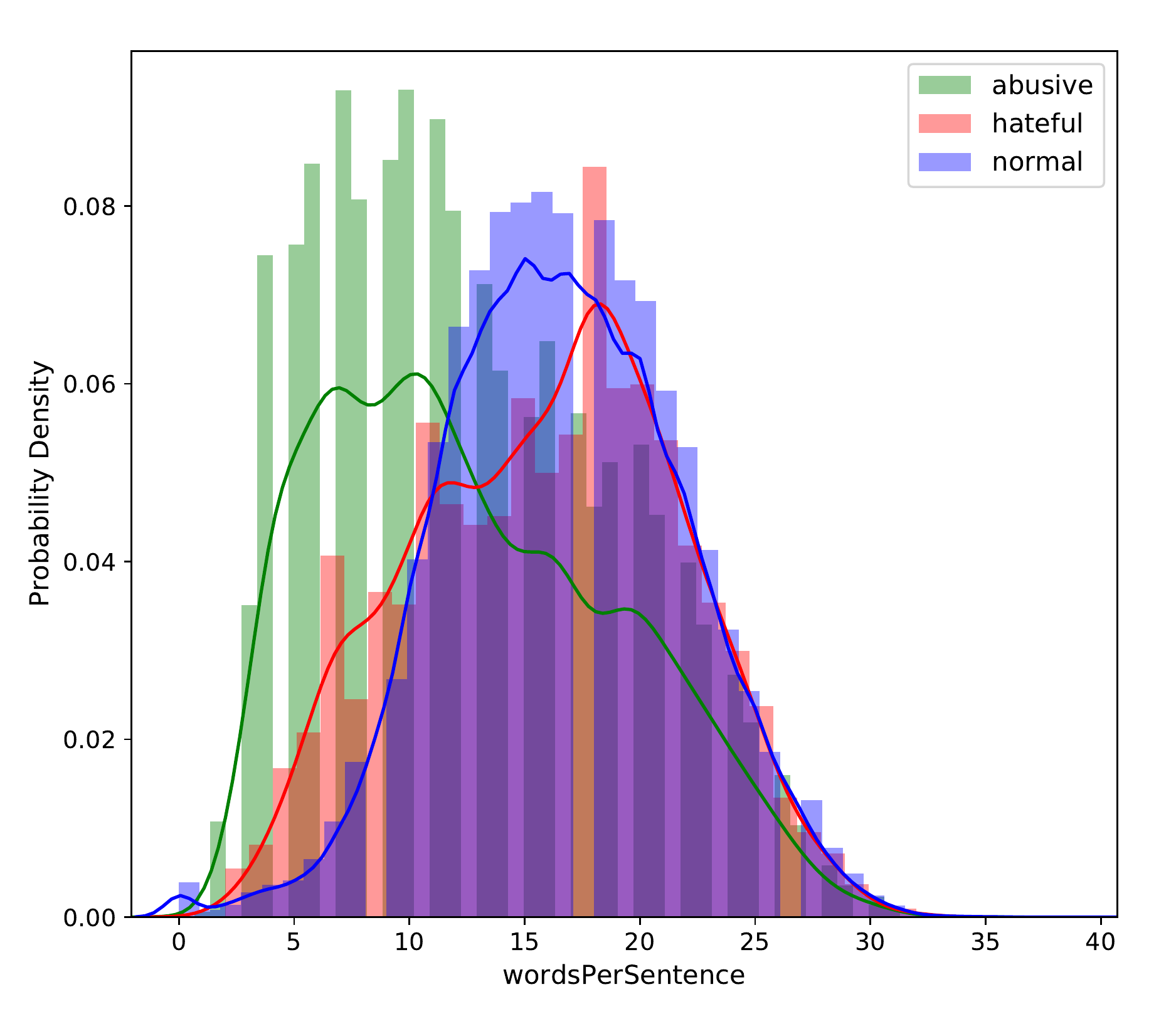}
		\caption{Mean words per sentence.}
		\label{fig:wordsPerSentence}
	\end{subfigure}
	\begin{subfigure}[b]{0.32\textwidth}
		\captionsetup{font=small}
		\includegraphics[width=\textwidth]{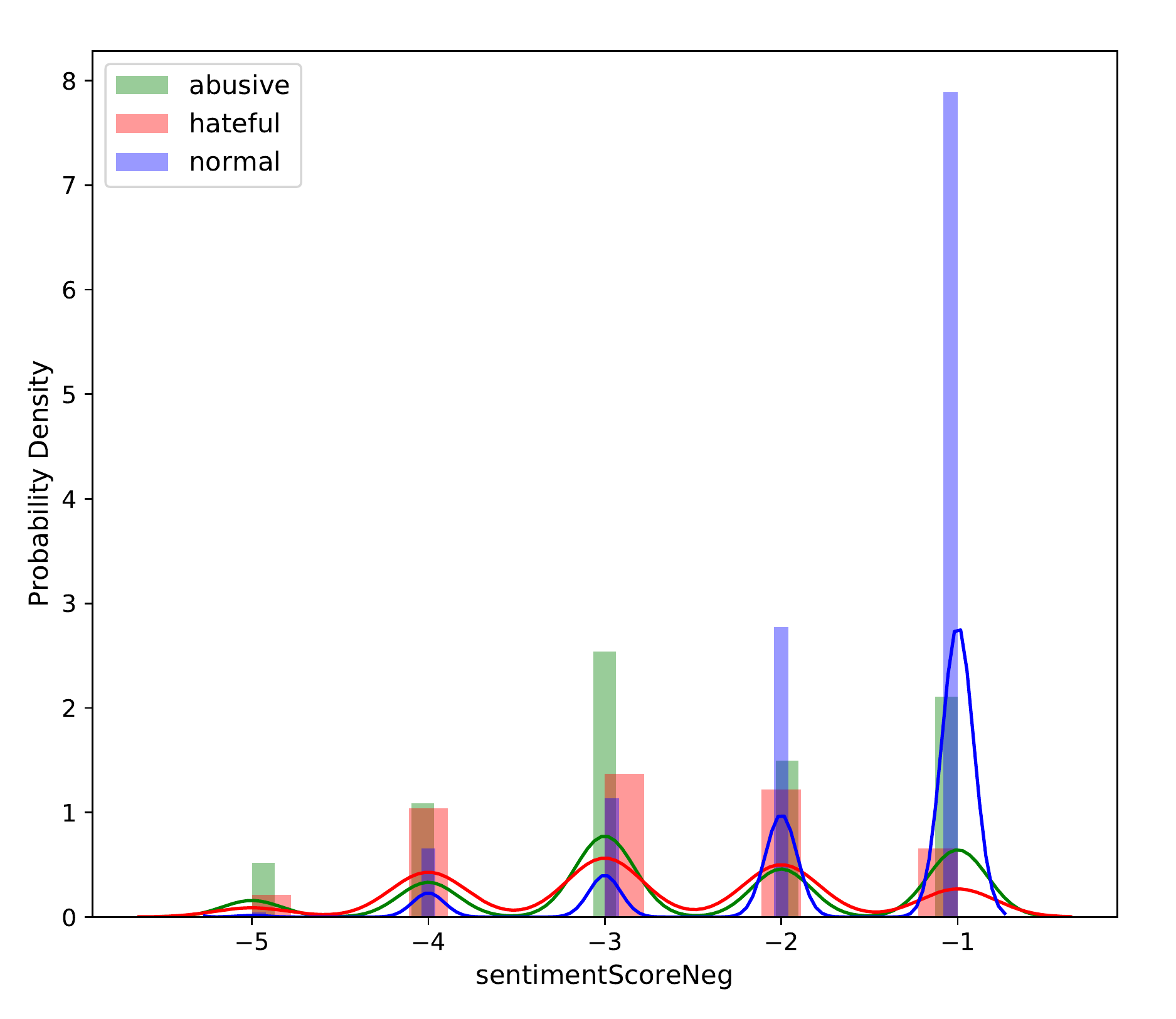}
		\caption{Sentiment (negative).}
		\label{fig:sentimentScoreNeg}
	\end{subfigure}	
	\begin{subfigure}[b]{0.32\textwidth}
		\captionsetup{font=small}
		\includegraphics[width=\textwidth]{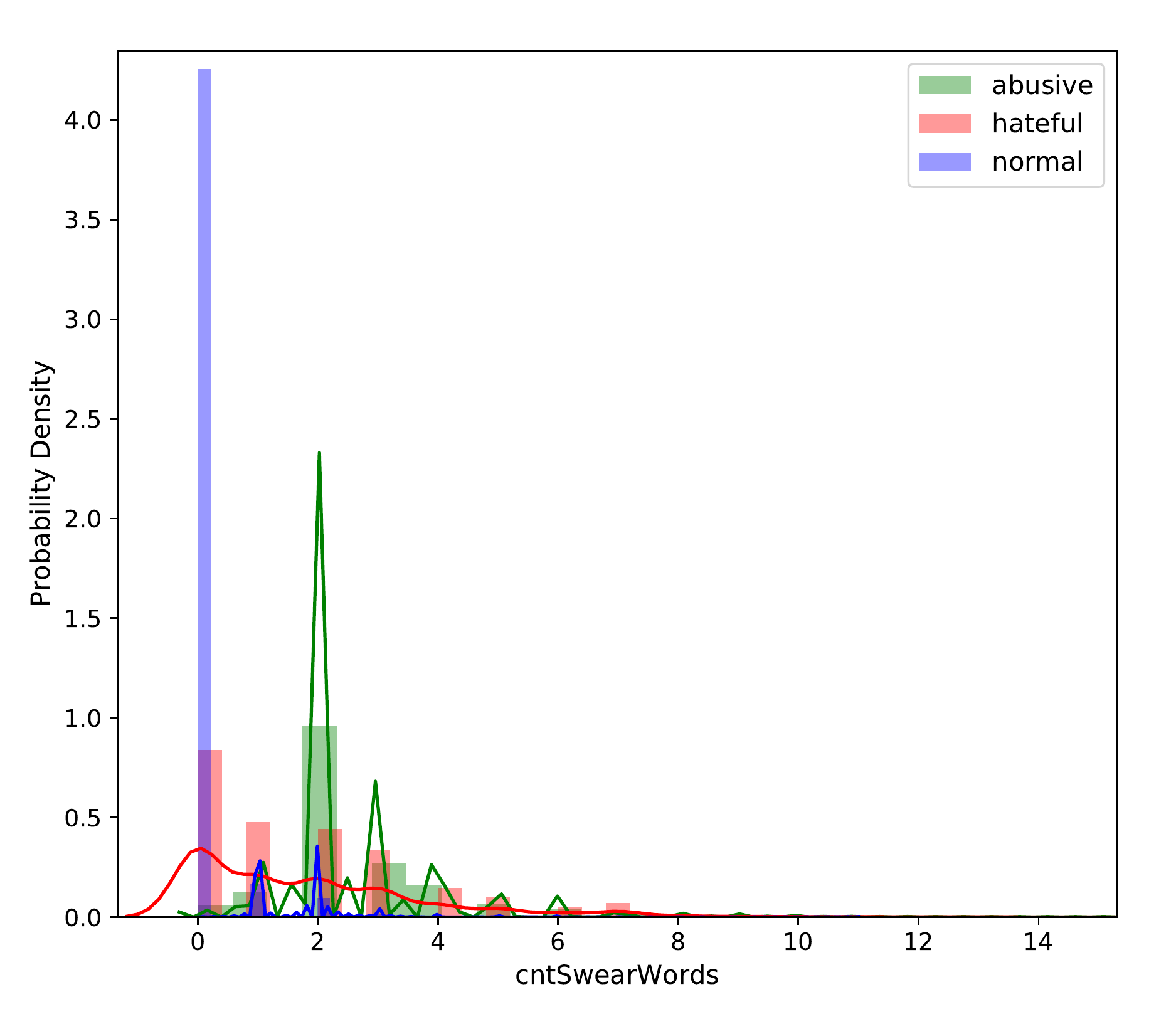}
		\caption{Swear words.}
		\label{fig:cntSwearWords}
	\end{subfigure}	
	\caption{PDF (Probability Density Function) of (a) Account age, (b) Uppercase letters, (c) Adjectives, (d) Mean words per sentence, (e) Sentiment (negative), and (f) Swear words.}
	\label{fig:features_distributions}
\end{figure*}

A wide range of characteristics can be considered for representing users' online presence and thus detecting the existence of abusive behavior.
Such features can come from either the users' profile, posted content, or their social network.
Authors in~\cite{chatzakou2019detecting} studied a wide range of features falling in the aforementioned categories. 
Inspired by their conducted analysis and after our own investigation, we have selected the most contributing features, with the aim of reducing execution time without sacrificing performance.
Indicatively, Figure~\ref{fig:features_distributions} plots the PDF (Probability Density Function) of some of the features presented next.

\descr{Profile Features.}
Features in this category include the age of the account (i.e., number of days since its creation),  the number of posts a user has made, and the number of lists subscribed to.
From Figure~\ref{fig:accountAge}, we observe that, on average, accounts that post normal content tend to be older ($1,487.74$ days) compared to those that post abusive ($1,291.97$ days) and hateful ($1,379.95$ days) content.

\descr{Text Features.}
This category includes basic text features, syntactic and stylistic ones, the sentiment expressed in the posted content, and the existence of swear words within posts.
\begin{itemize}
\item \textit{Basic:}  Number of hashtags used in a post, number of uppercase words, and number of URLs.
In Figure~\ref{fig:uppercases}, we plot the distribution of the number of uppercase words used in the abusive, hateful, and normal posts, finding that in the abusive and hateful posts, users tend to write with higher frequency in capital letters, which is quite an expected result since uppercase text can be indicative of intense emotional state or `shouting'.
Specifically, the average (STD) number of uppercase words for the normal, hateful, and abusive posts are $0.96$ ($2.10$), $1.57$ ($2.95$), and $1.84$ ($3.27$), respectively.

\item \textit{Syntactic:} We analyze the writing style of the authors of a text.
Specifically, we focus on Part-of-Speech (POS) features, which depict the relative frequency of adjectives, adverbs, and verbs in a text.
Figure~\ref{fig:adjectives} focuses on the number of adjectives used in the three studied categories of users' posts.
We observe that abusive and hateful posts tend to contain a lower number of adjectives in relation to the normal posts.
Adjectives, in general, are used to describe nouns and pronouns, and thus the lower use of them in the abusive/hateful posts indicates that users may prefer to carry out more direct attacks without adding additional information. 

\item \textit{Stylistic:} Mean number of words per sentence and mean word length.
Figure~\ref{fig:wordsPerSentence} depicts the PDF of the mean number of words per sentence.
It seems that normal posts follow a more similar pattern with the hateful posts, with normal posts containing somewhat more words ($16.66$ and $15.93$ words on average, for the normal and hateful posts, respectively).
As for the abusive posts, they tend to be shorter, and on average include $12.66$ words per sentence.

\item \textit{Sentiment:} In sentiment analysis, the objective is to determine the authors' attitude towards the topic under discussion. To estimate how positive or negative is the sentiment expressed in the posted content (on a $[-5, 5]$ scale), we utilized the SentiStrength tool~\cite{sentistrength}.
Comparing the distributions of the abusive and hateful classes with the normal for the negatively expressed sentiments (Figure~\ref{fig:sentimentScoreNeg}), we observe that normal posts include significantly less negative sentiments.

\item \textit{Swear Words:} To estimate the number of existing curse words within posts, we use a list of swear words obtained from~\cite{swearwords}.
Figure~\ref{fig:cntSwearWords} indicates that abusive and hateful posts contain a significantly higher number of swear words compared to the normal posts.
Specifically, the abusive and hateful posts include, on average, $2.54$ and $1.84$ swear words, respectively, where this number in case of the normal posts equals to $0.10$.
\end{itemize}

\smallskip
\noindent
Finally, in addition to the aforementioned features, we employ a new feature type, denoted as \textbf{adaptive Bag of Words (BoW)}.
For this feature, a BoW is initialized with swear words obtained from~\cite{swearwords} and is periodically enhanced based on tweet content.
Specifically, we maintain two sets of word counts and rolling statistics for aggressive (i.e., abusive and hateful) and normal tweets.
Words that occur frequently in aggressive tweets, but are not high-occurring words in normal tweets are added in the BoW.
Correspondingly, words that are becoming popular in normal tweets but lose traction in aggressive tweets are removed from the BoW.
Hence, this feature can adapt to changes in aggressive behaviors over time.

\descr{Network Features.}
This feature category aims to measure the popularity of a user based on two different criteria, i.e., the number of followers ({\it in-degree centrality}) and friends ({\it out-degree centrality}).
Overall, these measures can quantify a user's opportunity to have a positive or negative impact in their ego-network in a direct way.

\begin{figure}[!t]
	\centering
	\includegraphics[width=0.47\textwidth]{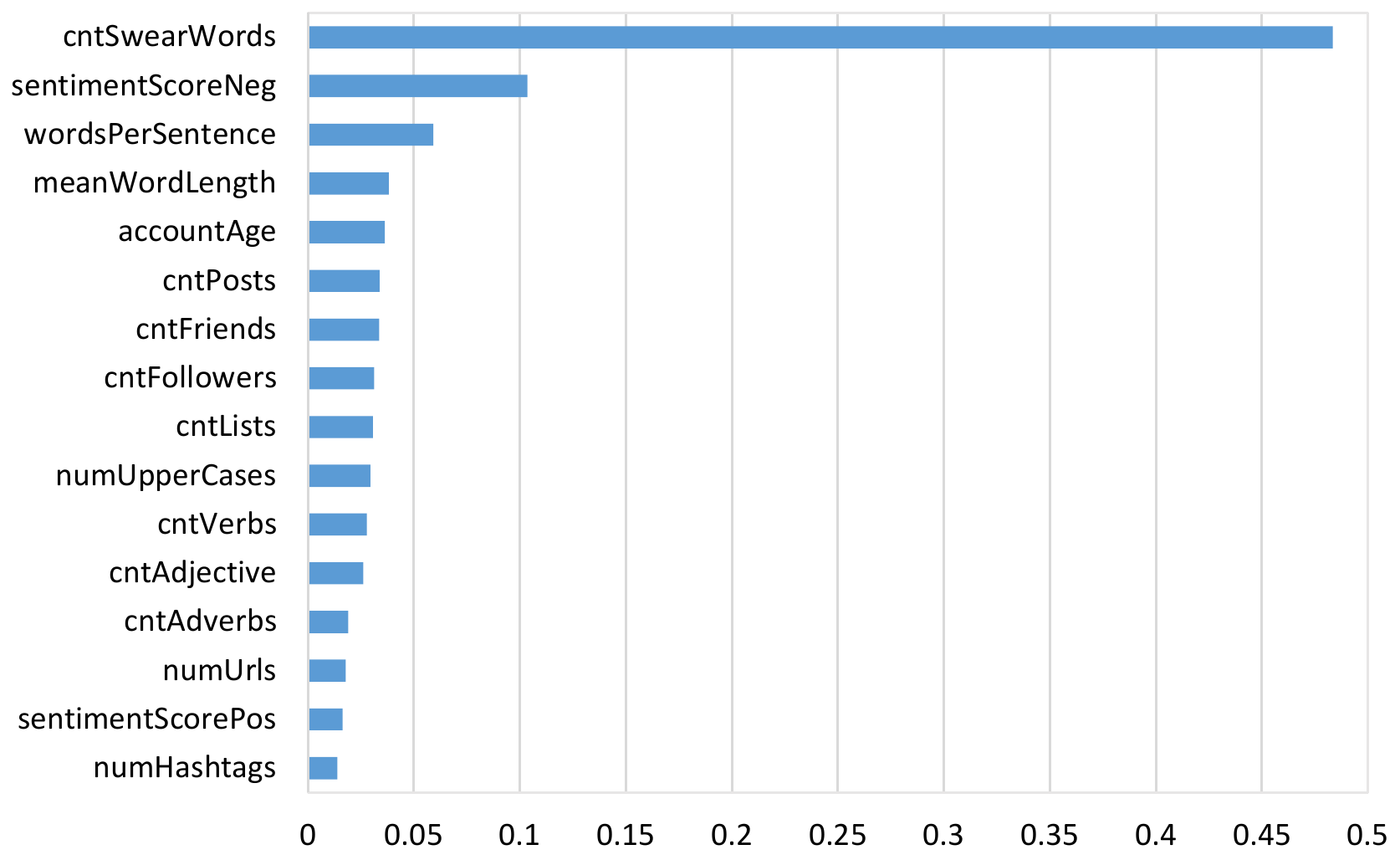}
	\caption{Evaluation of features based on $Gini$ importance.}
	\label{fig:features_eval}
\end{figure}

\descr{Features Evaluation.}
Figure~\ref{fig:features_eval} shows the feature relevance based on $Gini$ importance\footnote{The importance of a feature is computed as the (normalized) total reduction of the criterion brought by that feature.} in descending order.
We see that the most important feature is the swear count, followed by negative sentiment, number of words per sentence, mean word length, account age, and count posts.
Overall, we observe that the text-based features are among the most contributing ones.

%% file: sections/experiments.tex

\section{Experimental Evaluation}
\label{sec:experiments}

In this section, we evaluate the effectiveness and efficiency of our approach to correctly detect aggressive behavior in a large set of tweets.
Our evaluation methodology is as follows:
\begin{enumerate}
	\item We study the effect of preprocessing, normalization, and the adaptive BoW feature on the streaming ML performance (Section~\ref{sec:exp-preprocessing}).
	\item We evaluate the performance of various streaming ML methods (Section~\ref{sec:exp-streaming}).
	\item We compare the performance of streaming and batch-based methods (Section~\ref{sec:exp-streaming-vs-batch}).
	\item We investigate the capability of our method to scale and detect aggression on tweets at real-time (Section~\ref{sec:exp-scalability}).
	\item We explore how generalizable our approach is in detecting other related malign behaviors such as sarcasm, sexism, or racism (Section~\ref{sec:exp-other-datasets}).
\end{enumerate}

\subsection{Experimental Setup}
\label{sec:exp-setup}

Our experiments were executed on a server with an 8-core, 3.2GHz CPU, 64GB RAM, and a 2.1TB RAID 5 storage configuration.
To test the scalability of our system, we used a 3-node cluster, where each node has an 8-core, 2.4GHz CPU, 24GB RAM, and three 500GB hard drives.
All nodes were running CentOS Linux 7.2.
We implemented our approach over Apache Spark Streaming v2.3.2 and streamDM v0.2.

The dataset used consists of $86k$ tweets, from which $53,835$, $27,179$, and $4,970$ were labeled as \textit{normal}, \textit{abusive}, and \textit{hateful}, respectively.
Given the small number of \textit{hateful} tweets, we also experimented with the 2-class problem of detecting whether the tweets where \textit{normal} or \textit{aggressive} (i.e.,  \textit{abusive} or \textit{hateful}).

\descr{Hyperparameter Tuning.}
For each machine learning model (streaming or batch), we used grid search to find optimal parameter settings.
Table \ref{tab:hyperparam-tuning} lists the value ranges or options considered for each parameter for the three streaming ML models used in our experimental evaluation, namely Hoeffding Tree ($HT$), Adaptive Random Forest ($ARF$) of $HT$s, and Streaming Logistic Regression ($SLR$) with $SGD$.
The parameters and values for the equivalent batch models are similar.
Note that hyperparameter tuning in real-time is still an open-research problem for evolving data streams~\cite{gomes2019machine}.

\begin{table}[t]
	\caption{Hyperparameter tuning for streaming models}
	\label{tab:hyperparam-tuning}
	\begin{center}
	\begin{tabular}{llcc}
		\toprule
		Model & Parameter & Range or Options & Selected \\
		\midrule
		$HT$ & Split Criterion & Gini, InfoGain & InfoGain \\
		& Split Confidence & 0.001 - 0.5 & 0.01 \\
		& Tie Threshold & 0.01 - 0.1 & 0.05 \\
		& Grace Period & 200 - 500 & 200 \\
		& Max Tree Depth & 10 - 30 & 20 \\
		\midrule
		$ARF$ & \multicolumn{3}{l}{All $HT$ parameters above} \\
		& Ensemble Size & 10 - 20 & 10 \\
		\midrule
		$SLR$ & Lambda & 0.01 - 0.1 & 0.1 \\
		& Regularizer & Zero, L1, L2 & L2 \\
		& Regularization & 0.001 - 0.1 & 0.01 \\
		\bottomrule
	\end{tabular}
	\end{center}
\end{table}

\descr{Performance Metrics.}
In order to assess the performance of the ML methods, we measure typical ML metrics such as accuracy, precision, recall, and F1-score.
We picked \textit{F1-score} for demonstrating the performance of the methods, since it integrates both precision and recall measures into a unified measure, for a more general view of the performance.
For the streaming setting, we used the popular \textit{prequential evaluation} scheme, where instances are first used to test, and then to train the streaming ML model.

\subsection{Effect of Preprocessing, Normalization, and Adaptive BoW}
\label{sec:exp-preprocessing}

\begin{figure}[t]
\begin{center}
	\includegraphics[width=\columnwidth]{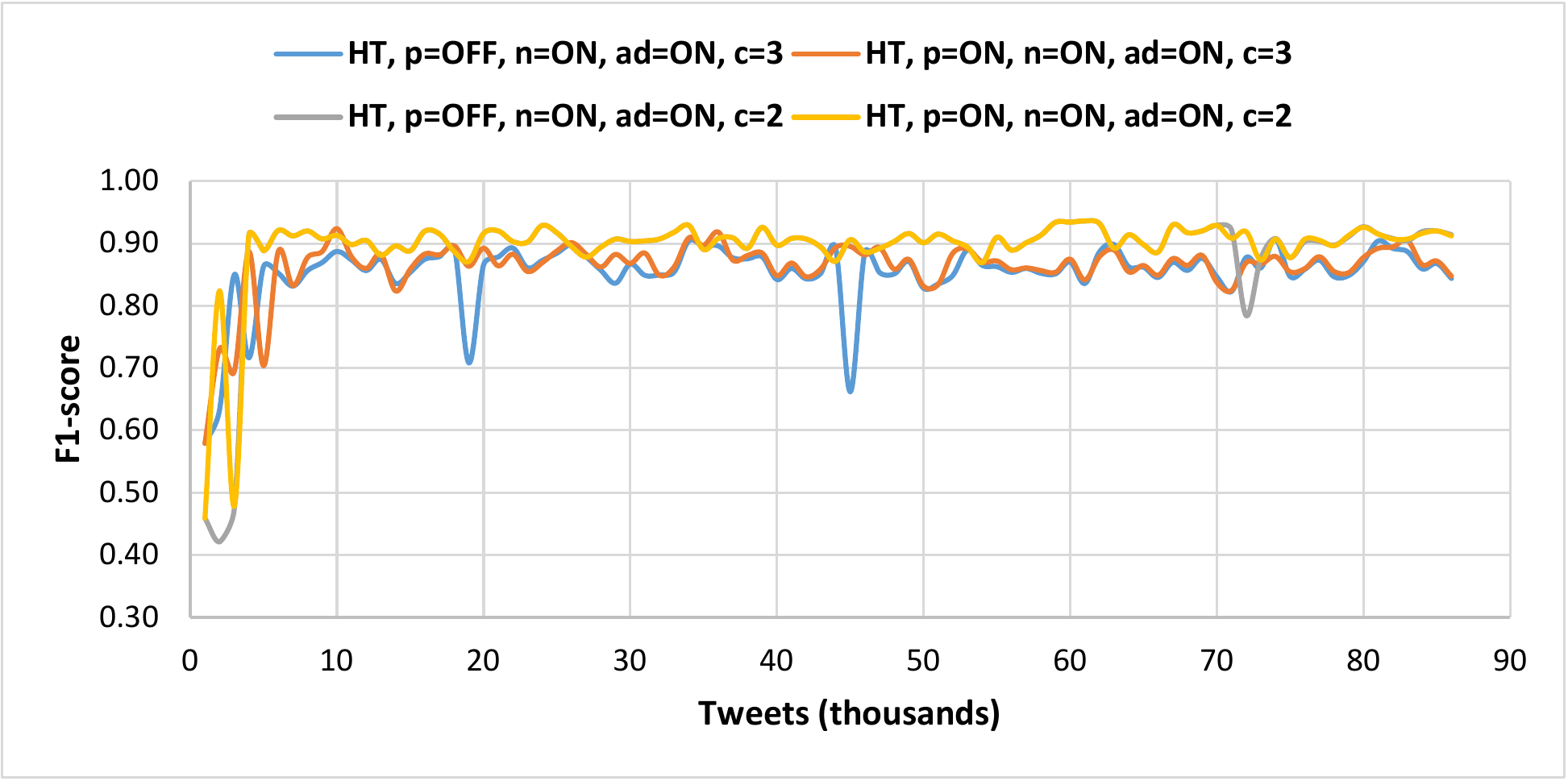}
\caption{F1-score for $HT$ for 2- or 3-class problem (c=2 or 3), when we apply normalization (n=ON) and adaptive bag of words (ad=ON), and test the impact of preprocessing (p=ON or OFF).}
\label{fig:ht-pX-nON-cX-aON-f1}
\end{center}
\end{figure}

\begin{figure}[t]
\begin{center}
	\includegraphics[width=\columnwidth]{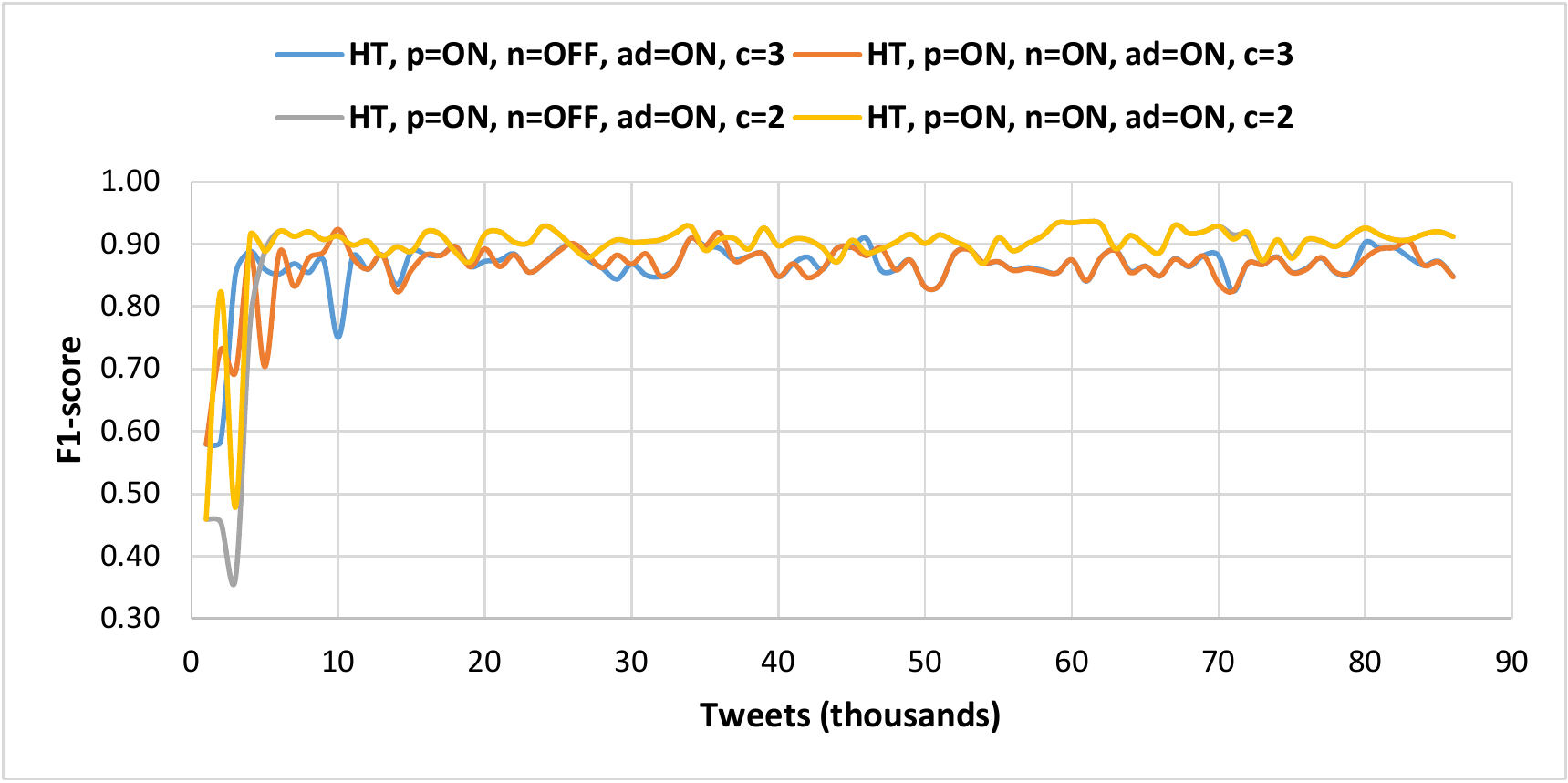}
\caption{F1-score for $HT$ for 2- or 3-class problem (c=2 or 3), when we apply preprocessing (p=ON) and adaptive bag of words (ad=ON), and test the impact of normalization (n=ON or OFF).}
\label{fig:ht-pON-nX-cX-aON-f1}
\end{center}
\end{figure}

\begin{figure}[t]
\begin{center}
	\includegraphics[width=\columnwidth]{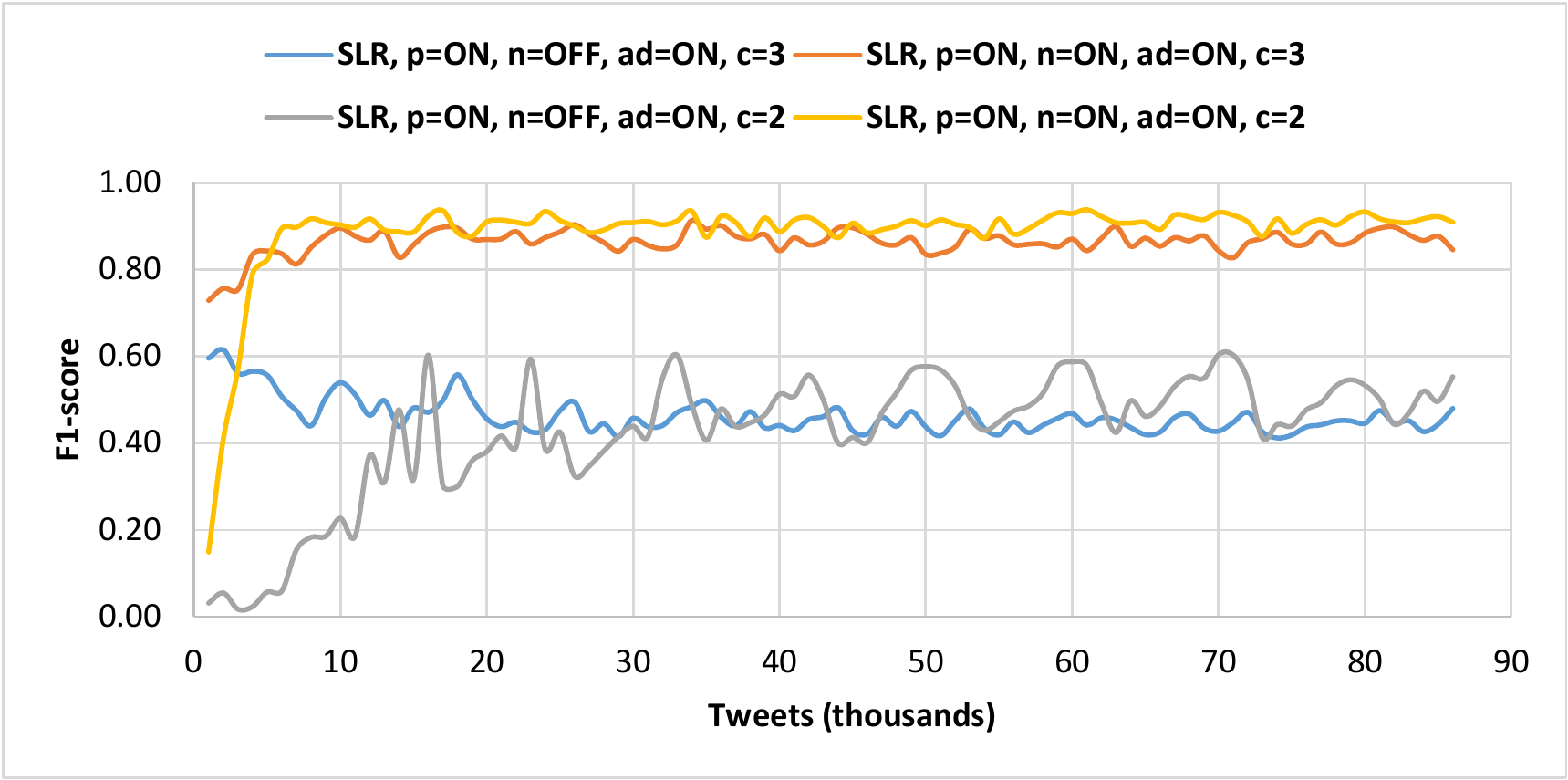}
\caption{F1-score for $SLR$ for 2- or 3-class problem (c=2 or 3), when we apply preprocessing (p=ON) and adaptive bag of words (ad=ON), and test the impact of normalization (n=ON or OFF).}
\label{fig:rf-pON-nX-cX-aON-f1}
\end{center}
\end{figure}

\begin{figure}[t]
\begin{center}
	\includegraphics[width=\columnwidth]{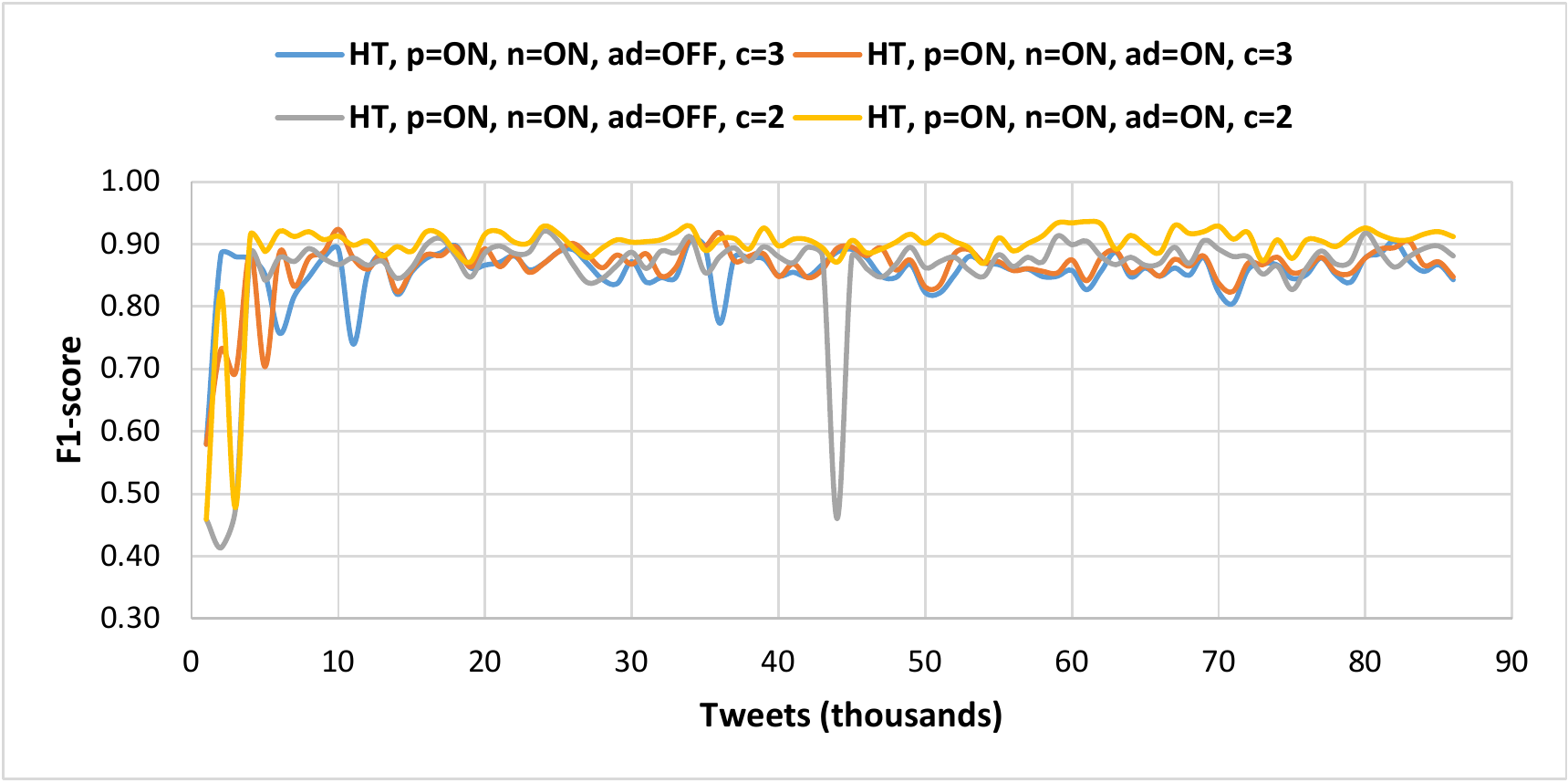}
\caption{F1-score for $HT$ for 2- or 3-class problem (c=2 or 3), when we apply preprocessing (p=ON) and normalization (n=ON), and test the impact of applying adaptive bag of words (ad=ON or OFF).}
\label{fig:ht-pON-nON-cX-aX-f1}
\end{center}
\end{figure}

\descr{Preprocessing.}
First, we investigate the impact of preprocessing on the performance of streaming ML methods.
We only show results for $HT$ as they are similar with the other two methods.
In Figure~\ref{fig:ht-pX-nON-cX-aON-f1} we compare the F1-score of $HT$ with and without preprocessing of tweets.
This is done for both the 2- and 3-class problems while applying normalization and the adaptive bag-of-words feature.
We find that preprocessing helps the classifier's performance, by smoothing out transient changes in the data and allowing the classifier to perform in a more stable fashion.
We also notice that when binary classification is performed, the performance is better than the 3-class problem.
In the 3-class case, the small number of \textit{hateful} tweets makes it harder to correctly classify that label, reducing the overall F1-score performance by about $4\%$ compared to the 2-class case.

\descr{Normalization.}
Next, we study the impact that normalization has to the streaming ML methods.
We tested the three forms of normalization described in Section \ref{sec:system:overview} and found that \textit{minmax without outliers} leads to a slightly better performance (around $2\%$).
Hence, this is the form we used for our subsequent experiments.
Figure~\ref{fig:ht-pON-nX-cX-aON-f1} reveals that enabling or disabling normalization has a marginal effect on the performance of $HT$ for both the 2- and 3-class problems.
However, the same is not true for the other two methods.
As Figure~\ref{fig:rf-pON-nX-cX-aON-f1} demonstrates, enabling normalization in $SLR$ increases F1-score by over $42\%$, regardless of class setup, while it also smooths out performance.
For $ARF$, the impact of normalization is also profound and there can be up to $30$-$60\%$ difference in F1-score for the two class setups (not shown due to space constrains).

\descr{Adaptive BoW.}
In Figure~\ref{fig:ht-pON-nON-cX-aX-f1}, we demonstrate the results when applying the adaptive bag-of-words feature in $HT$ as opposed to using a fixed bag-of-words (recall Section~\ref{sec:datasets:fextraction}).
We find that employing this feature helps the ML performance with an average improvement of $2$-$4\%$ on the F1-score; this is true for both the 2- and 3-class setup.
We also note that the adaptive list helps the ML performance to be smoother and more resilient in changes of word distribution through time.
These results are similar for the other two streaming ML methods.
The size of the adaptive list fluctuates through time due to the addition and removal of swear words as the word distributions in normal and aggressive tweets change.
However, the overall size of the list does not greatly increase through time.
For example, this list is initialized with $347$ swear words obtained from~\cite{swearwords}, and after the $86k$ tweets are parsed, it reaches $529$ words.

\descr{Key takeaways.}
Preprocessing and adaptive BoW offer modest benefits on performance (about $2$-$4\%$ higher F1-score) and better stability through time.
Normalization, on the other hand, is model-dependent: it has a negligible impact for $HT$ but a significant impact on $ARF$ and $SLR$.
Finally, all methods perform better on the 2-class, compared to the 3-class problem.

\subsection{Performance of Streaming ML Methods}
\label{sec:exp-streaming}

After investigating the preliminary results on how preprocessing, normalization, and adaptive BoW affect performance of the detection methods, we enable all three and proceed to compare results for all streaming ML methods tested, for the 3- and 2-class problems.
Table \ref{tab:eval-metrics} shows some key evaluation metrics for the three methods for the 3- and 2-class problems.

\begin{table}[t]
	\caption{Key evaluation metrics for $HT$, $ARF$, and $SLR$}
	\label{tab:eval-metrics}
	\begin{center}
		\begin{tabular}{l@{~~~~~}ccc@{~~~~~}ccc}
			\toprule
			Metric & \multicolumn{3}{c@{~~~~~}}{3-class} & \multicolumn{3}{c}{2-class} \\
			& HT & ARF & SLR & HT & ARF & SLR \\
			\midrule
			Accuracy & 0.89 & 0.85 & 0.89 & 0.93 & 0.92 & 0.93 \\
			Precision & 0.85 & 0.80 & 0.85 & 0.92 & 0.85 & 0.91 \\
			Recall & 0.89 & 0.85 & 0.89 & 0.90 & 0.93 & 0.91 \\
			F1-score & 0.87 & 0.83 & 0.87 & 0.91 & 0.89 & 0.91 \\
			\bottomrule
		\end{tabular}
	\end{center}
\end{table}

\begin{figure}[t]
	\begin{center}
		\includegraphics[width=\columnwidth]{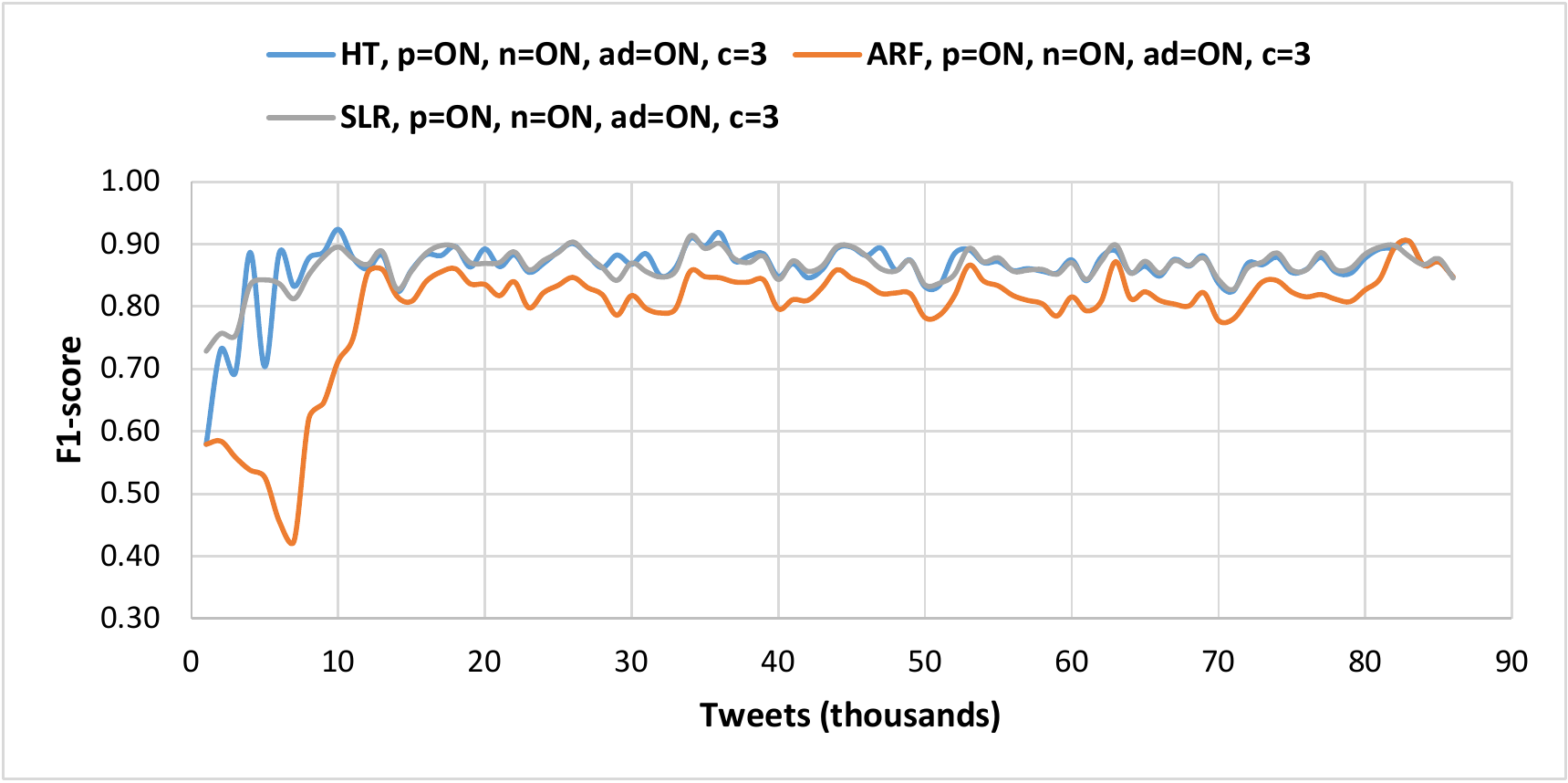}
		\caption{F1-score for $HT$, $ARF$, and $SLR$ for 3-class problem (c=3), when we apply preprocessing, normalization and adaptive bag of words (p=ON, n=ON, ad=ON).}
		\label{fig:ht-arf-slr-pON-nON-c3-aON-f1.pdf}
	\end{center}
\end{figure}

\descr{3-class Problem.}
Figure~\ref{fig:ht-arf-slr-pON-nON-c3-aON-f1.pdf} shows results for the 3-class problem.
We notice that all methods achieve an F1-score in the range of $80$-$90\%$.
The $HT$ and $SLR$ offer a similar high performance, with $HT$ occasionally reaching higher F1-scores ($0.5\%$ higher on average).
On the other hand, $ARF$ always underperforms with about $4\%$ less F1-score over time.
Through time, all methods exhibit similar trends, i.e., their variability in performance is similar.
This means they are affected in a similar manner by transient changes in the distributions of the available features.
In addition, $HT$ and $SLR$ take $\sim$$5$-$9k$ instances to reach their full capacity with respect to detection performance.
Interestingly, $ARF$ takes longer ($\sim$$12k$ instances) to plateau its performance.

\begin{figure}[t]
\begin{center}
	\includegraphics[width=\columnwidth]{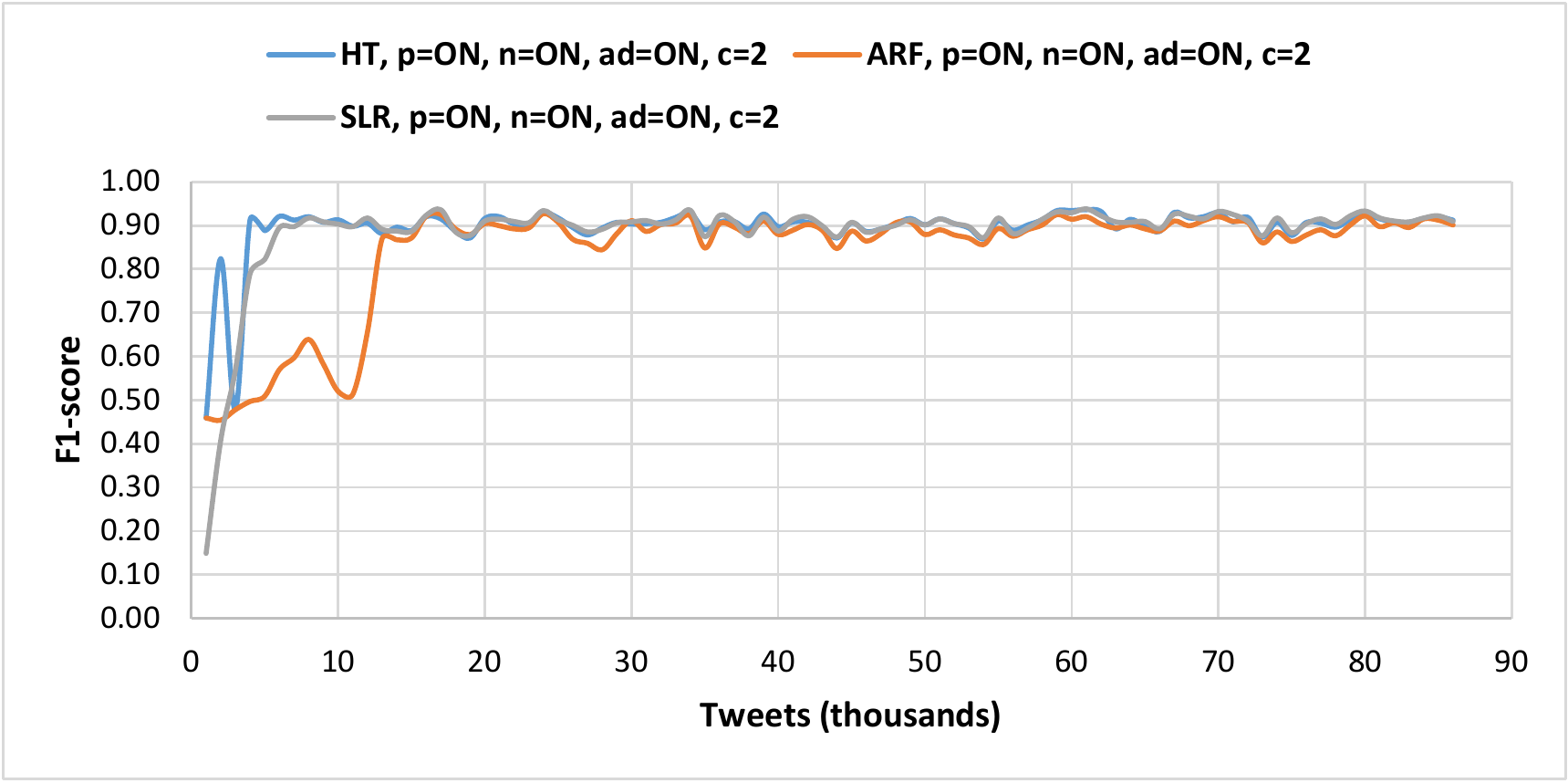}
\caption{F1-score for $HT$, $ARF$, and $SLR$ for 2-class problem (c=2), when we apply preprocessing, normalization and adaptive bag of words (p=ON, n=ON, ad=ON).}
\label{fig:ht-arf-slr-pON-nON-c2-aON-f1}
\end{center}
\end{figure}

\descr{2-class Problem.}
In a similar fashion, Figure~\ref{fig:ht-arf-slr-pON-nON-c2-aON-f1} shows the results for the 2-class problem.
This is a simpler classification problem, and thus, the ML performance increases across all methods and reaches to over $91\%$ F1-score.
In fact, $HT$'s performance is increased by up to $4\%$ in the F1-score.
This is expected given that the minority \textit{hateful} label is now merged with the \textit{abusive} label and the data imbalance is avoided.
Finally, $HT$ and $SLR$ are able to reach their full potential quicker than the 3-class problem (after $\sim$$4k$ tweets), whereas $ARF$ requires around the same number of tweets ($\sim$$12k$).

\descr{Key takeaways.}
$HT$ either performs better or similar to the other streaming ML methods (with $87$-$91\%$ F1-score) and it reaches its peak performance quickly (after $4$-$6k$ tweets).

\subsection{Streaming vs. Batch ML Methods}
\label{sec:exp-streaming-vs-batch}

One of the key differences between streaming and batch ML methods is that streaming methods only observe each instance once, whereas batch methods typically process each instance multiple times.
Hence, it is important to compare the performance of our three streaming methods with corresponding (or similar) batch methods.
In particular, we tested the Decision Tree $J48$, Random Forest, and Logistic Regression using the ML software $WEKA$ v3.7~\cite{weka}.

As our Twitter dataset was collected over a period of 10 consecutive days (of $\sim$$8$-$9k$ tweets each day), we examine the following two training scenarios for our batch methods.
In the ``train-first-day\_test-all-others'', the batch method is trained in the data of the first day, and then it is only tested on the data of each subsequent day.
This scenario represents the case where the model is not updated, and can become stale over time.
The second scenario is the ``train-one-day\_test-next-day'', where the batch method is trained on data of one day and tested on the next day. 
Then, it is trained on the data of the subsequent day and tested on the day after, etc.
This scenario represents a case closer to what a pseudo-streaming method would be doing: periodically (daily in this case) update the model on newly acquired and labeled data.

\begin{figure}[t]
	\begin{center}
		\includegraphics[width=\columnwidth]{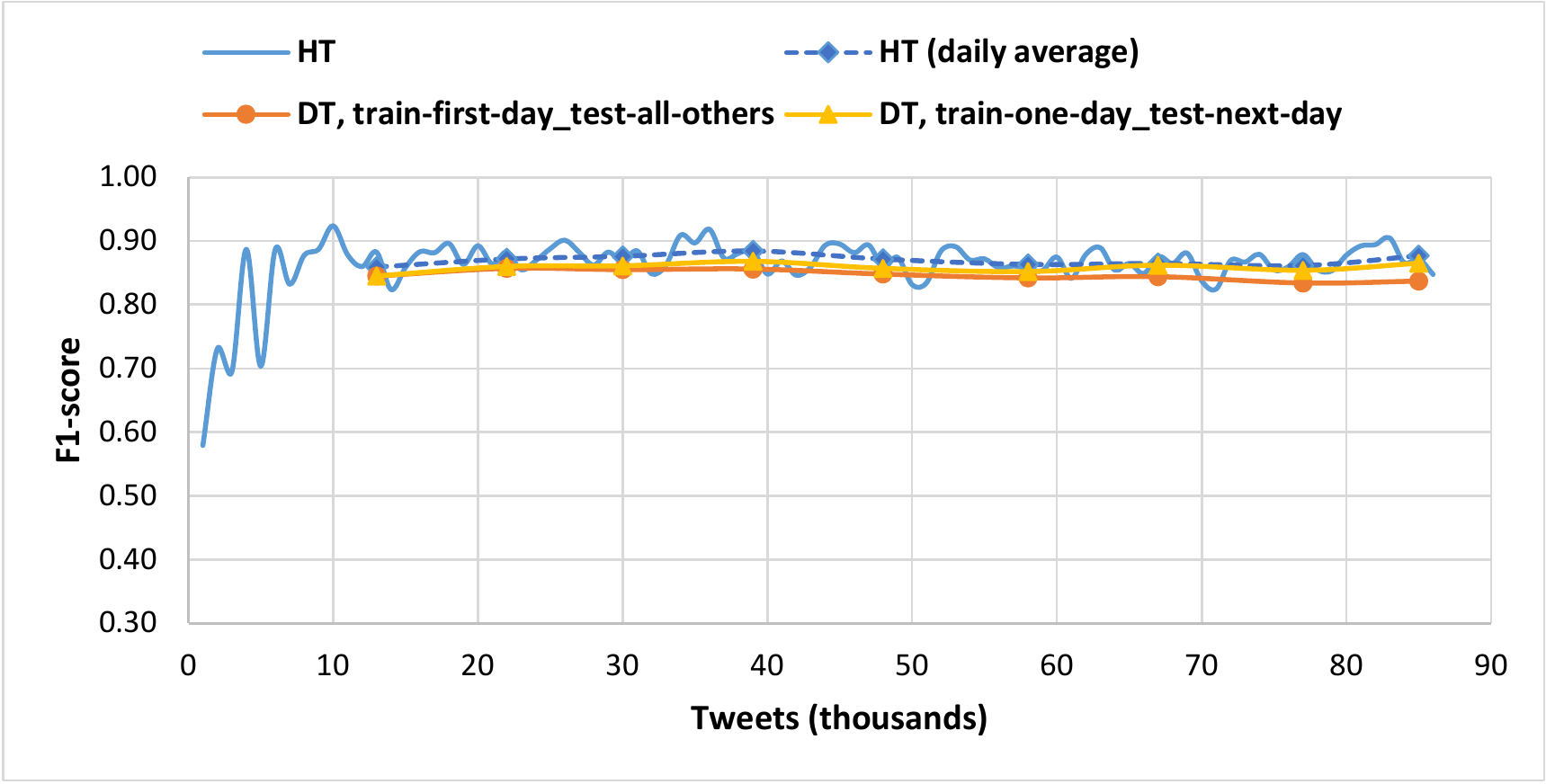}
		\caption{F1-score for $HT$ vs. $DT$, for 3-class problem, for two batch-based training methods (train-first-day\_test-all-others and train-one-day\_test-next-day).}
		\label{fig:ht-vs-batch-f1-3class}
	\end{center}
\end{figure}

\begin{figure}[t]
	\begin{center}
		\includegraphics[width=\columnwidth]{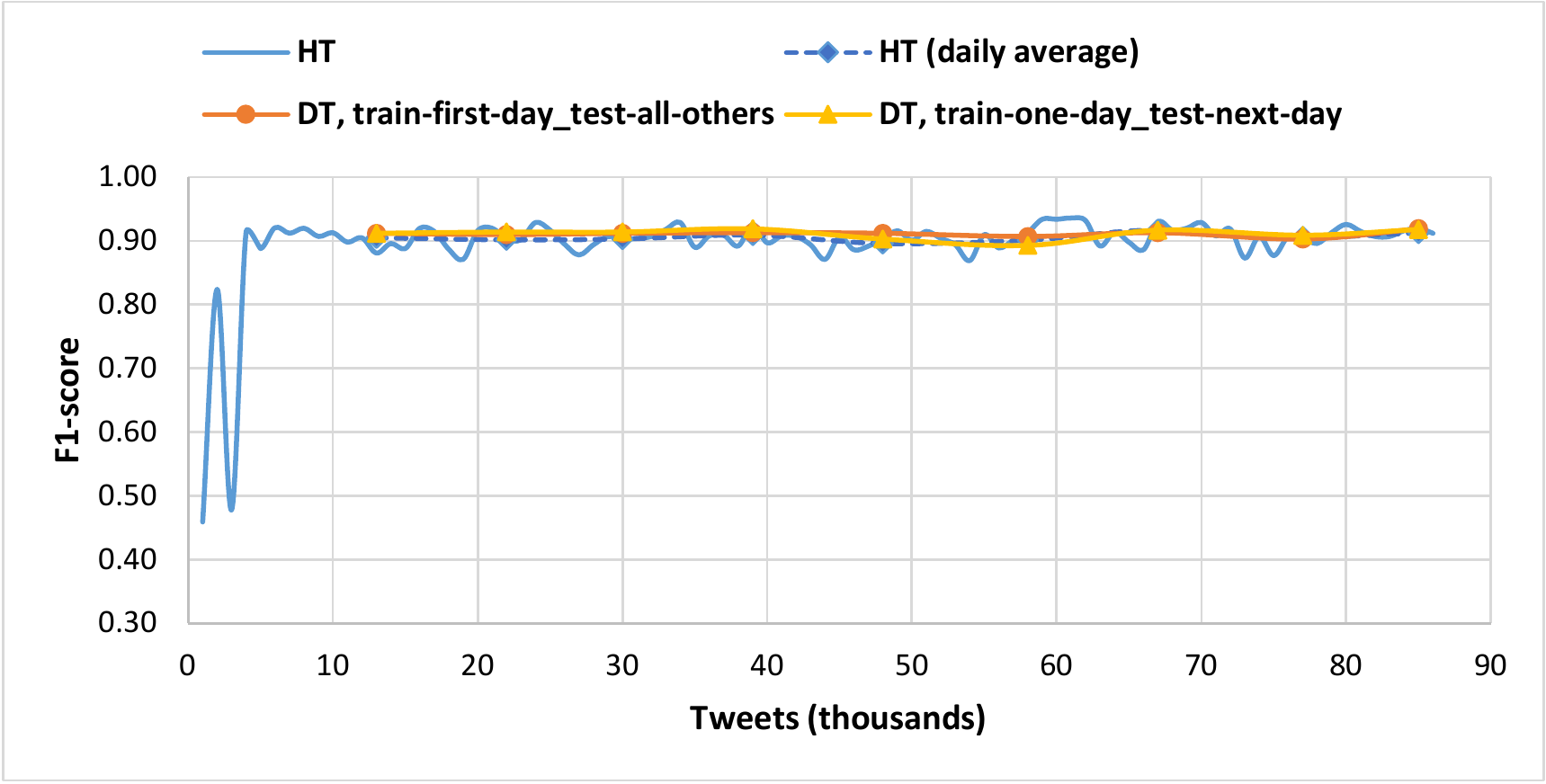}
		\caption{F1-score for $HT$ vs. $DT$, for 2-class problem, for two batch-based training methods (train-first-day\_test-all-others and train-one-day\_test-next-day).}
		\label{fig:ht-vs-batch-f1-2class}
		\vspace{-7pt}
	\end{center}
\end{figure}

Interestingly, the performance of the batch methods were very similar in all scenarios.
Thus, we only show comparison results between the best streaming method ($HT$) and its corresponding batch method, i.e., the decision tree ($DT$).

\descr{3-class Problem.}
Under the 3-class problem, the $HT$ performs at least as good as the batch method ($DT$) for both training scenarios, as shown in Figure~\ref{fig:ht-vs-batch-f1-3class}.
Interestingly, and expectedly, the performance of the batch method for the first scenario (``train-first-day\_test-all-others'') degrades slowly through time.
After the processing of $86k$ tweets, we observe a degradation of about $2\%$ in F1-score from the beginning of the experiment, when the training was finalized on the 1st day.
In the second setup, the batch method performs somewhat better: the model gets updated with new data every day, and therefore, it can keep up with transient changes in the features' distributions.
However, $HT$ still performs better than this batch method for this training scenario, as it is able to adapt to changes in a more fine-grained manner.

\descr{2-class Problem.}
In Figure~\ref{fig:ht-vs-batch-f1-2class}, we repeat the same experiment but for the 2-class problem.
We observe that the performance of the batch method ($DT$) for the two training scenarios is very similar and fairly stable through time.
As for the $HT$'s performance, it is slightly lower the first few days compared to the $DT$, with less than $1\%$ difference on average in F1-score.
Afterwards, the $HT$'s performance is on par with the $DT$'s performance.

\descr{Key takeaways.}
Even though a streaming ML method processes each instance only once, it is able to achieve the same performance -- and in some cases even better -- compared to similar, batch-based ML methods.

\subsection{Scalability for Real-time Detection on Twitter}
\label{sec:exp-scalability}

\begin{figure}[t]
\begin{center}
	\includegraphics[width=\columnwidth]{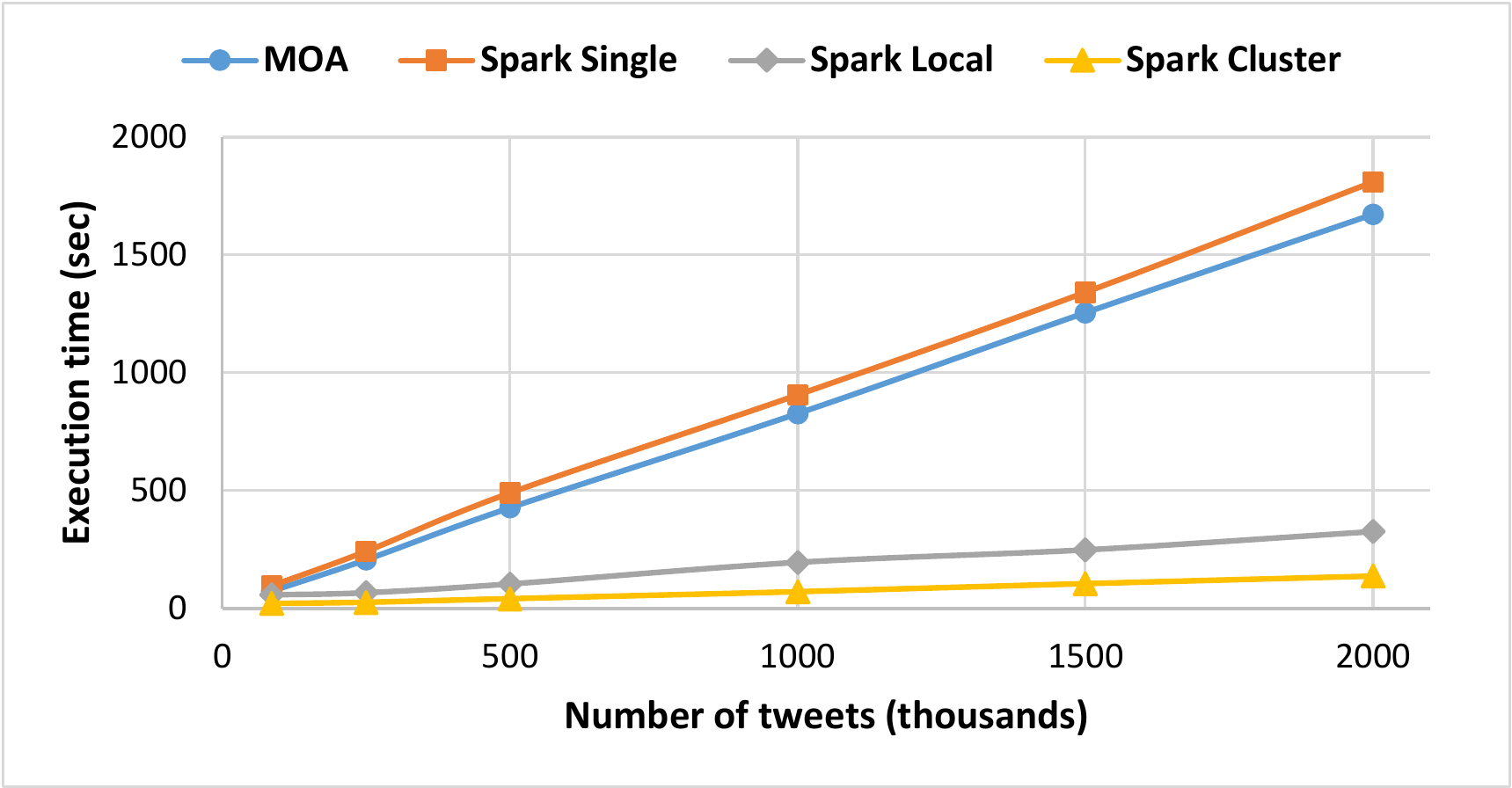}
	\caption{Execution time per streaming system, given a number of incoming tweets to be processed.}
	\label{fig:exec-time-vs-tweets-different-systems}
\end{center}
\end{figure}

In this section, we study the capability for our framework to gracefully scale and process more tweets through time.
For our evaluation, we test different flavors of $Spark$ execution, namely single-threaded ($Spark Single$), multi-threaded on one machine ($Spark Local$), and multi-threaded across 3 machines ($Spark Cluster$), with 8 threads per machine.
For comparison purposes, we also implemented our approach over $MOA$ v19.05.0, an open-source framework containing several state-of-the-art ML algorithms for data streams~\cite{moa-website}.
However, $MOA$ is a single-threaded ML engine and does not allow parallelized processing.

For each setting, we process a fixed number of unlabeled tweets (ranged from $250k$ to $2m$ tweets) intermixed with the $86k$ labeled tweets.
We measure the computation cost in terms of total execution time, which accounts for the entire processing pipelining (feature extraction, training, prediction, and evaluation; recall Section~\ref{sec:system:overview}).
Finally, we also study throughput under each setting, defined as the total number of tweets processed per second.
The results presented next pertain to using $HT$ with preprocessing, normalization, and the adaptive BoW feature enabled for the 3-class problem.

\begin{figure}[t]
	\begin{center}
	\includegraphics[width=\columnwidth]{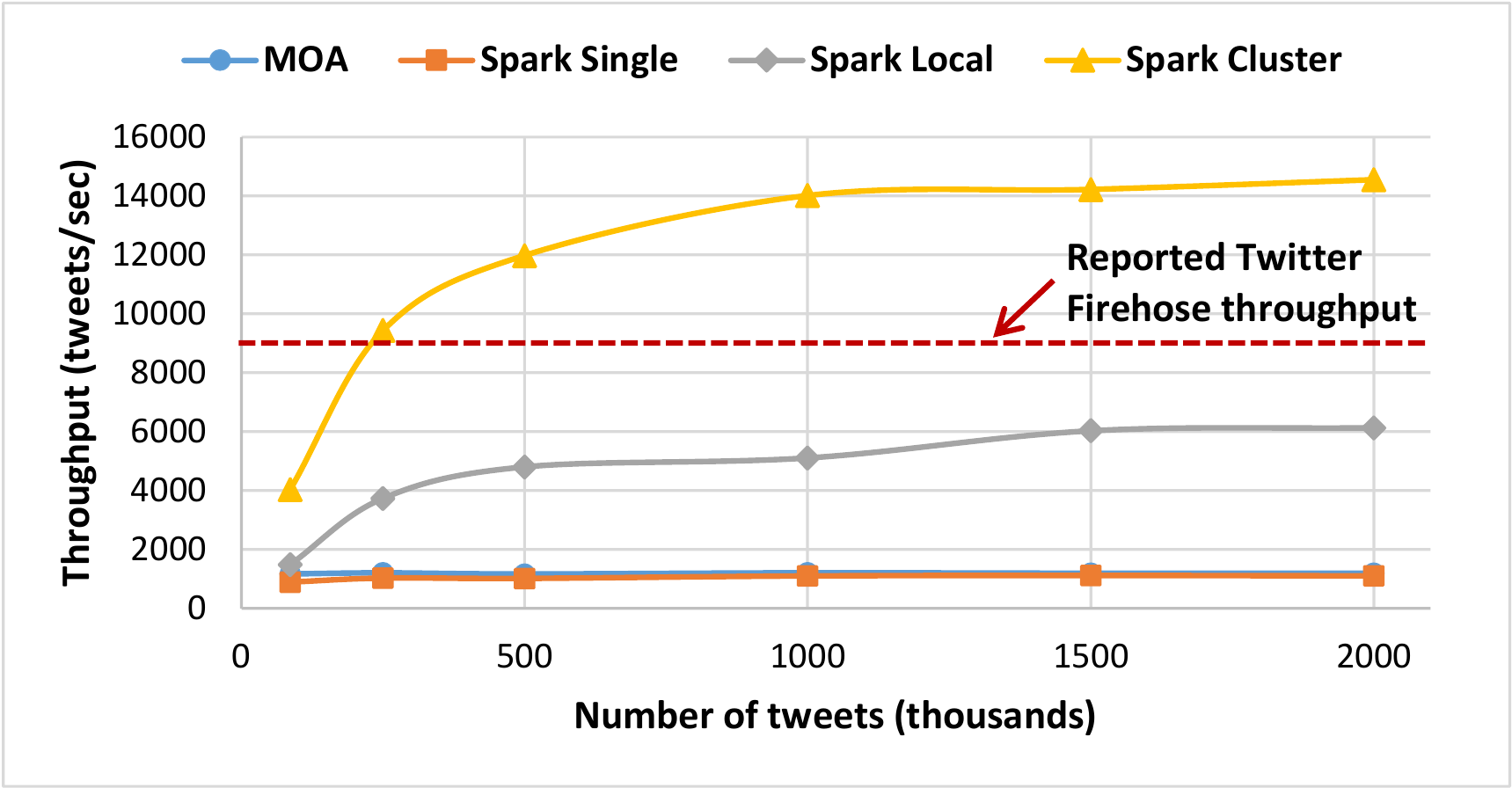}
	\caption{Throughput per streaming system, given a number of incoming tweets to be processed.}
	\label{fig:throughput-vs-tweets-different-systems}
	\vspace{-6pt}
	\end{center}
\end{figure}

First, we compare how the execution time varies for processing a fixed number of tweets, while using $MOA$ or $Spark$ as the underlying framework (Figure~\ref{fig:exec-time-vs-tweets-different-systems}).
On the one hand, we confirm that $MOA$ and $Spark Single$ have a similar performance, where doubling the number of tweets to be processed approximately doubles the execution time taken to process them.
This is to be expected, since the two frameworks operate on one machine and process tweets sequentially.
Interestingly, $Spark Single$ exhibits about $7$-$17\%$ worse performance than $MOA$ due to management overheads in $Spark$ engine for scheduling the micro-batches.
This is again expected, as $Spark$ has been designed to optimally process big data in parallel infrastructures, instead of sequentially such as in $MOA$.

On the other hand, we find that when $Spark$ is allowed to operate in a multi-threaded environment, on a single or multiple machines, the performance greatly improves.
Such a setup, can process the same number of tweets much faster than the single-threaded $Spark$ or $MOA$.
For example, to process $2$ million tweets, $Spark Local$ and $Spark Cluster$ needed $5.5\times$ and $13.2\times$ less time, respectively, than $Spark Single$.
Also, $Spark Cluster$ processed the same number of tweets in $2.5\times$ less time than $Spark Local$.

Finally, we compare the throughput of these framework setups in tweets per second (Figure~\ref{fig:throughput-vs-tweets-different-systems}).
We clearly see major difference in performance of $Spark Cluster$ over $Spark Local$, and the superiority of such a setup over the single-threaded $Spark$ and $MOA$ (which have a constant throughput of $\sim$$1100$ tweets per seconds).
In particular, we observe that with a single worker, the multi-threaded $Spark Local$ setup can process up to $6k$ tweets per second.
When the machines triple, the $Spark Cluster$ can process up to $14.5k$ tweets per second.
Interestingly, these two setups demonstrate a plateau in their throughput performance after about 1 million tweets are fed into the ML pipeline.

\descr{Key takeaways.}
These performance results clearly demonstrate the scalability of our streaming approach, in order to achieve real-time aggression detection.
In fact, if we assume the publicly claimed number of arrivals of tweets in the overall Twitter Firehose to be about $9k$ tweets per second~\cite{internet-live-stats}, our results show that the proposed ML pipeline could easily tackle this stream of tweets and detect such aggressive behavior with only $3$ Spark machines.

\subsection{Detecting Related Behaviors in Real Time}
\label{sec:exp-other-datasets}

\begin{figure}[t]
	\begin{center}
		\includegraphics[width=\columnwidth]{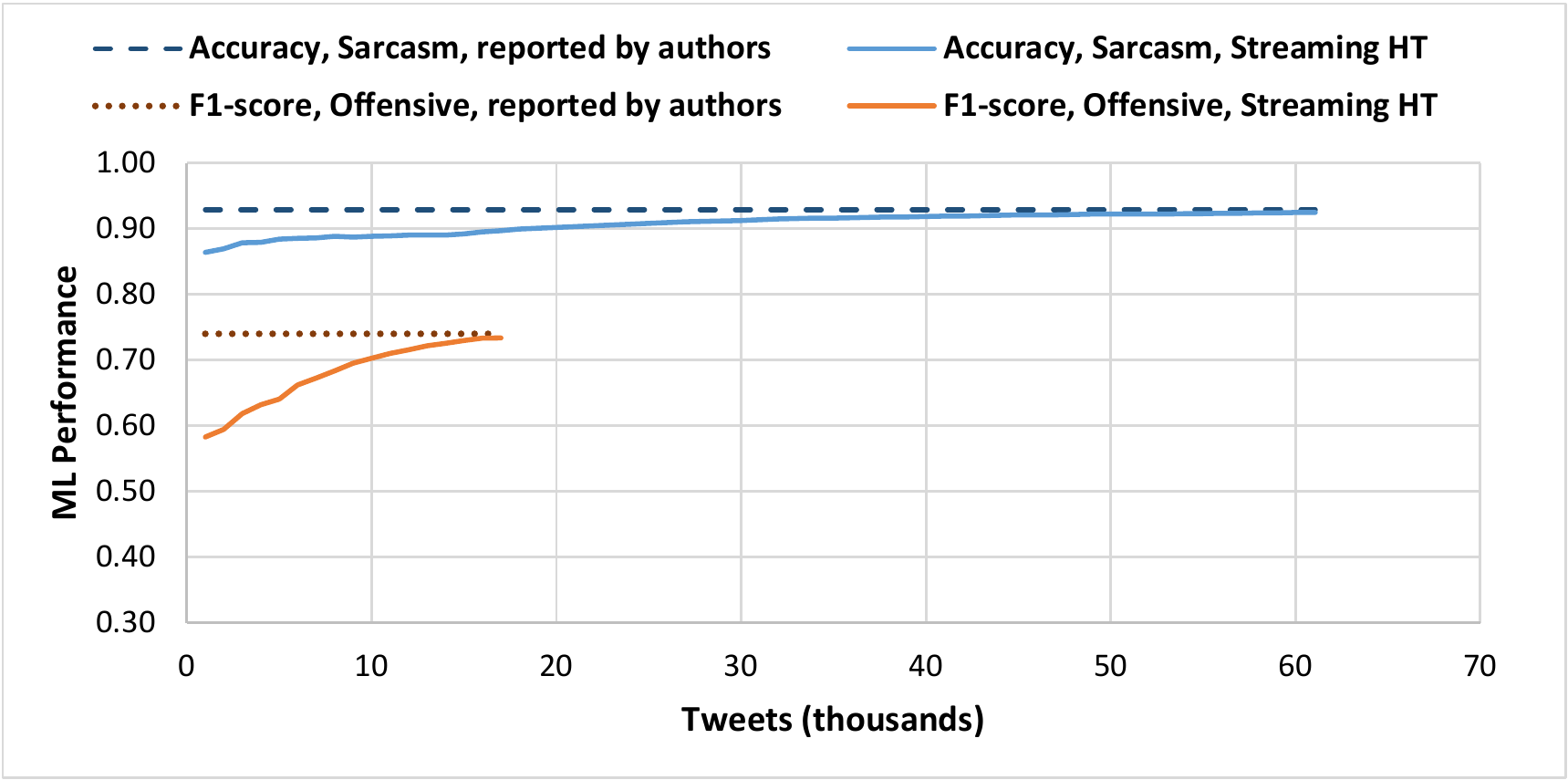}
		\caption{ML performance of $HT$ vs. original (batch-based) performance for the Sarcasm~\cite{rajadesingan2015sarcasm} and Offensive~\cite{waseem2016hateful} datasets.}
		\label{fig:ml-perf-other-datasets}
		\vspace{-2pt}
	\end{center}
\end{figure}

This section investigates whether our streaming ML approach can be adopted to detect other undesired behaviors on Twitter, in addition to aggressive behavior.
For his purpose, we acquired two more datasets:
\begin{itemize}
	\item The \textit{Sarcasm} dataset is focused on detecting sarcasm in tweets, and is provided by~\cite{rajadesingan2015sarcasm}. The dataset contains $6.5k$ sarcastic tweets, out of $61k$ total.
	\item The \textit{Offensive} dataset is focused on detecting racism and sexism in tweets, and is provided by~\cite{waseem2016hateful}. The dataset contains $2k$ racist and $3k$ sexists tweets, out of $16k$ total.
\end{itemize}
We tested both datasets using the $HT$ model, trained and evaluated with the original features as described in the respective papers.
Figure~\ref{fig:ml-perf-other-datasets} compares the performance achieved by $HT$ with the best performance reported by the authors.
Both datasets were originally evaluated using logistic regression and 10-fold cross validation.
For the \textit{Sarcasm} dataset, an accuracy of $93\%$ was reported in~\cite{rajadesingan2015sarcasm} (F1-score was not available).
The $HT$ model started with an accuracy of $86\%$, which gradually increased through time. Accuracy crossed the $90\%$ mark after $19k$ tweets and slowly converged towards $93\%$.
For the Offensive dataset, a $74\%$ F1-score was reported in~\cite{waseem2016hateful}.
Even though the $HT$'s performance started at $58\%$ F1-score, it quickly increased to $73\%$ after processing the $16k$ tweets.

\descr{Key takeaways.}
The streaming ML paradigm can be applied for detecting other malign behaviors in Twitter, with its performance reaching similar levels over time as the batch-based ML methods.

%% file: sections/conclusion.tex

\section{Conclusion and Future Work}
\label{sec:conclusion}

In this work, we have developed an aggression detection framework for social media, where the entire processing pipeline -- from preprocessing and feature extraction to training and testing -- employs the data streaming paradigm.
As a result, the streaming ML model is incrementally updated and remains always up-to-date, despite any transient aggression behaviors.
At the same time, the framework is remarkably scalable, as it is capable to consume and potentially process the entire Twitter Firehose with only $3$ commodity machines.
The evaluation results show that the streaming ML models can perform better or similar to popular batch-based ML methods, achieving over $90\%$ accuracy, precision, and recall after processing just a few thousand tweets.
Finally, we showed that our proposed streaming ML framework can be easily applied to detect other types of abusive behavior such as sarcasm, racism, and sexism, with minimal effort.

As part of future work, we plan to extend our current prototype into a unified framework for the real-time detection of a variety of behaviors (e.g., bullying, aggressive, abusive, offensive, racist, sexist) for diverse social media platforms such as Twitter, Facebook, Instagram, YouTube, and Snapchat.
As some forms of behaviors, like cyberbullying and trolling, usually involve repetitive hostile actions, we also plan to investigate detecting such behaviors at the level of media sessions (e.g., for a group of tweets from the same user, for a set of comments under a post).
For this purpose, we can utilize the windowing functionalities provided by all distributed stream processing engines for performing the groupings in a scalable and timely manner.

%% file: paper.bbl
\begin{thebibliography}{10}
\providecommand{\url}[1]{#1}
\csname url@samestyle\endcsname
\providecommand{\newblock}{\relax}
\providecommand{\bibinfo}[2]{#2}
\providecommand{\BIBentrySTDinterwordspacing}{\spaceskip=0pt\relax}
\providecommand{\BIBentryALTinterwordstretchfactor}{4}
\providecommand{\BIBentryALTinterwordspacing}{\spaceskip=\fontdimen2\font plus
\BIBentryALTinterwordstretchfactor\fontdimen3\font minus
  \fontdimen4\font\relax}
\providecommand{\BIBforeignlanguage}[2]{{%
\expandafter\ifx\csname l@#1\endcsname\relax
\typeout{** WARNING: IEEEtran.bst: No hyphenation pattern has been}%
\typeout{** loaded for the language `#1'. Using the pattern for}%
\typeout{** the default language instead.}%
\else
\language=\csname l@#1\endcsname
\fi
#2}}
\providecommand{\BIBdecl}{\relax}
\BIBdecl

\bibitem{cyberbullying-facts}
{Cyberbullying Research Center}, \url{https://cyberbullying.org/facts}, 2019.

\bibitem{passive-aggression}
S.~Whitson, ``{Confronting Passive Aggressive Behavior on Social Media},''
  \url{http://bit.ly/2QP1yi9}, May 2018.

\bibitem{dinakar2011modeling}
K.~Dinakar, R.~Reichart, and H.~Lieberman, ``{Modeling the Detection of Textual
  Cyberbullying},'' \emph{The Social Mobile Web}, vol.~11, no.~2, 2011.

\bibitem{meanBirds}
D.~Chatzakou, N.~Kourtellis, J.~Blackburn, E.~De~Cristofaro, G.~Stringhini, and
  A.~Vakali, ``{Mean Birds: Detecting Aggression and Bullying on Twitter},'' in
  \emph{WebSci}.\hskip 1em plus 0.5em minus 0.4em\relax ACM, 2017, pp. 13--22.

\bibitem{chatzakou2019detecting}
D.~Chatzakou, I.~Leontiadis, J.~Blackburn, E.~D. Cristofaro, G.~Stringhini,
  A.~Vakali, and N.~Kourtellis, ``{Detecting Cyberbullying and Cyberaggression
  in Social Media},'' \emph{ACM Transactions on the Web (TWEB)}, vol.~13,
  no.~3, pp. 1--51, 2019.

\bibitem{founta2018large}
A.~M. Founta, C.~Djouvas, D.~Chatzakou, I.~Leontiadis, J.~Blackburn,
  G.~Stringhini, A.~Vakali, M.~Sirivianos, and N.~Kourtellis, ``{Large Scale
  Crowdsourcing and Characterization of Twitter Abusive Behavior},'' in
  \emph{ICWSM}.\hskip 1em plus 0.5em minus 0.4em\relax AAAI, 2018.

\bibitem{riadi2017detection}
Hariani and I.~Riadi, ``{Detection Of Cyberbullying On Social Media Using Data
  Mining Techniques},'' \emph{IJCSIS}, vol.~15, no.~3, p. 244, 2017.

\bibitem{waseem2016hateful}
Z.~Waseem and D.~Hovy, ``{Hateful Symbols or Hateful People? Predictive
  Features for Hate Speech Detection on Twitter},'' in \emph{SRW@ HLT-NAACL},
  2016, pp. 88--93.

\bibitem{davidson2017automated}
T.~Davidson, D.~Warmsley, M.~Macy, and I.~Weber, ``{Automated Hate Speech
  Detection and the Problem of Offensive Language},'' in \emph{ICWSM}.\hskip
  1em plus 0.5em minus 0.4em\relax AAAI, 2017.

\bibitem{lee2002holistic}
J.~Lee and Y.~Lee, ``{A Holistic Model of Computer Abuse within
  Organizations},'' \emph{Information management \& computer security},
  vol.~10, no.~2, pp. 57--63, 2002.

\bibitem{Chen2012DetectingOffensiveLanguage}
Y.~Chen, Y.~Zhou, S.~Zhu, and H.~Xu, ``{Detecting Offensive Language in Social
  Media to Protect Adolescent Online Safety},'' in \emph{{PASSAT and
  SocialCom}}, 2012.

\bibitem{nobata2016abusive}
C.~Nobata, J.~Tetreault, A.~Thomas, Y.~Mehdad, and Y.~Chang, ``{Abusive
  Language Detection in Online User Content},'' in \emph{WWW}.\hskip 1em plus
  0.5em minus 0.4em\relax ACM, 2016, pp. 145--153.

\bibitem{Hosseinmardi2015}
H.~Hosseinmardi, S.~A. Mattson, R.~I. Rafiq, R.~Han, Q.~Lv, and S.~Mishra,
  ``{Analyzing Labeled Cyberbullying Incidents on the Instagram Social
  Network},'' in \emph{{SocInfo}}, 2015.

\bibitem{yao2019cyberbullying}
M.~Yao, C.~Chelmis, and D.~S. Zois, ``{Cyberbullying Ends Here: Towards Robust
  Detection of Cyberbullying in Social Media},'' in \emph{WWW}.\hskip 1em plus
  0.5em minus 0.4em\relax ACM, 2019, pp. 3427--3433.

\bibitem{Djuric2015HateSpeechDetection}
N.~Djuric, J.~Zhou, R.~Morris, M.~Grbovic, V.~Radosavljevic, and
  N.~Bhamidipati, ``{Hate Speech Detection with Comment Embeddings},'' in
  \emph{{WWW}}.\hskip 1em plus 0.5em minus 0.4em\relax ACM, 2015, pp. 29--30.

\bibitem{kayes2015ya-abuse}
I.~Kayes, N.~Kourtellis, D.~Quercia, and F.~Iamnitchi, A. \&~Bonchi, ``{The
  Social World of Content Abusers in Community Question Answering},'' in
  \emph{{WWW}}.\hskip 1em plus 0.5em minus 0.4em\relax ACM, 2015.

\bibitem{4chan}
G.~E. Hine, J.~Onaolapo, E.~De~Cristofaro, N.~Kourtellis, I.~Leontiadis,
  R.~Samaras, G.~Stringhini, and J.~Blackburn, ``{Kek, Cucks, and God Emperor
  Trump: A Measurement Study of 4chan's Politically Incorrect Forum and Its
  Effects on the Web},'' in \emph{{ICWSM}}.\hskip 1em plus 0.5em minus
  0.4em\relax AAAI, 2017.

\bibitem{Smith2008CyberbullyingNature}
P.~Smith, J.~Mahdavi, M.~Carvalho, S.~Fisher, S.~Russell, and N.~Tippett,
  ``{Cyberbullying: Its Nature and Impact in Secondary School Pupils},''
  \emph{{Child Psychology and Psychiatry}}, vol.~49, no.~4, pp. 376--385, 2008.

\bibitem{bbc2020Twitter-bullying-feature}
BBC, ``{Twitter to Test 'block all replies' Function},'' January {2020},
  \url{https://www.bbc.com/news/business-51044086}.

\bibitem{instagram2019Bullying-feature}
Instagram, ``{Introducing the ``Restrict'' Feature to Protect Against
  Bullying},'' October {2019},
  \url{https://about.instagram.com/blog/announcements/stand-up-against-bullying-with-restrict}.

\bibitem{social-media-stats}
J.~Bagadiya, ``{309 Social Media Statistics You Must Know In 2020},''
  \url{https://www.socialpilot.co/blog/social-media-statistics}, {2020}.

\bibitem{internet-live-stats}
``{Internet Live Stats in 1 Second},''
  \url{https://www.internetlivestats.com/one-second/}, {2020}.

\bibitem{social-norms-chi18}
K.~R. Allison, ``{Social Norms in Online Communities: Formation, Evolution and
  Relation to Cyber-Aggression},'' in \emph{Extended Abstracts of CHI}.\hskip
  1em plus 0.5em minus 0.4em\relax ACM, 2018, p. 1–4.

\bibitem{rafiq2018scalable}
R.~I. Rafiq, H.~Hosseinmardi, R.~Han, Q.~Lv, and S.~Mishra, ``{Scalable and
  Timely Detection of Cyberbullying in Online Social Networks},'' in
  \emph{SCA}.\hskip 1em plus 0.5em minus 0.4em\relax ACM, 2018, pp. 1738--1747.

\bibitem{zhang2018detecting}
Z.~Zhang, D.~Robinson, and J.~Tepper, ``{Detecting Hate Speech on Twitter Using
  a Convolution-GRU Based Deep Neural Network},'' in \emph{European Semantic
  Web Conference}.\hskip 1em plus 0.5em minus 0.4em\relax Springer, 2018, pp.
  745--760.

\bibitem{badjatiya2017deep}
P.~Badjatiya, S.~Gupta, M.~Gupta, and V.~Varma, ``{Deep Learning for Hate
  Speech Detection in Tweets},'' in \emph{WWW}.\hskip 1em plus 0.5em minus
  0.4em\relax ACM, 2017, pp. 759--760.

\bibitem{warner2012detecting}
W.~Warner and J.~Hirschberg, ``{Detecting Hate Speech on the World Wide Web},''
  in \emph{2nd Workshop on Language in Social Media}, 2012, pp. 19--26.

\bibitem{burnap2015cyber}
P.~Burnap and M.~L. Williams, ``{Cyber Hate Speech on Twitter: An Application
  of Machine Classification and Statistical Modeling for Policy and Decision
  Making},'' \emph{Policy \& Internet}, vol.~7, no.~2, pp. 223--242, 2015.

\bibitem{joulin2016bag}
A.~Joulin, E.~Grave, P.~Bojanowski, and T.~Mikolov, ``{Bag of Tricks for
  Efficient Text Classification},'' \emph{arXiv preprint arXiv:1607.01759},
  2016.

\bibitem{park2017one}
J.~H. Park and P.~Fung, ``{One-step and Two-step Classification for Abusive
  Language Detection on Twitter},'' \emph{arXiv preprint arXiv:1706.01206},
  2017.

\bibitem{herodotou2021icde-streaming-aggression}
H.~Herodotou, D.~Chatzakou, and N.~Kourtellis, ``{Catching them red-handed:
  Real-time Aggression Detection on Social Media},'' in \emph{{ICDE}}.\hskip
  1em plus 0.5em minus 0.4em\relax IEEE, 2021.

\bibitem{spark-streaming}
``{\em Apache Spark Streaming},'' \url{https://spark.apache.org/streaming/},
  2020.

\bibitem{spark-rdd-nsdi12}
M.~Zaharia, M.~Chowdhury, T.~Das, A.~Dave, J.~Ma, M.~McCauley, M.~J. Franklin,
  S.~Shenker, and I.~Stoica, ``{Resilient Distributed Datasets: A
  Fault-tolerant Abstraction for In-memory Cluster Computing},'' in
  \emph{NSDI}.\hskip 1em plus 0.5em minus 0.4em\relax {USENIX} Association,
  2012, pp. 2--14.

\bibitem{moa-jmlr10}
A.~Bifet, G.~Holmes, R.~Kirkby, and B.~Pfahringer, ``{MOA: Massive Online
  Analysis},'' \emph{Journal of Machine Learning Research}, vol.~11, no. May,
  pp. 1601--1604, 2010.

\bibitem{samoa-website}
``{\em Apache Samoa},'' \url{https://samoa.incubator.apache.org/}, 2020.

\bibitem{streamdm}
``{\em streamDM: Data Mining for Spark Streaming},''
  \url{http://huawei-noah.github.io/streamDM/}, 2020.

\bibitem{hoeffding-tree-kdd00}
P.~Domingos and G.~Hulten, ``{Mining High-speed Data Streams},'' in
  \emph{SIGKDD}.\hskip 1em plus 0.5em minus 0.4em\relax ACM, 2000, pp. 71--80.

\bibitem{arf-ml17}
H.~M. Gomes, A.~Bifet, J.~Read, J.~P. Barddal, F.~Enembreck, B.~Pfharinger,
  G.~Holmes, and T.~Abdessalem, ``{Adaptive Random Forests for Evolving Data
  Stream Classification},'' \emph{Machine Learning}, vol. 106, no. 9-10, pp.
  1469--1495, 2017.

\bibitem{giatsoglou2015retweeting}
M.~Giatsoglou, D.~Chatzakou, N.~Shah, C.~Faloutsos, and A.~Vakali,
  ``{Retweeting Activity on Twitter: Signs of Deception},'' in
  \emph{PAKDD}.\hskip 1em plus 0.5em minus 0.4em\relax Springer, 2015, pp.
  122--134.

\bibitem{sentistrength}
{SentiStrength}, \url{http://sentistrength.wlv.ac.uk/}, {2017}.

\bibitem{swearwords}
AllSlang, ``List of swear words \& curse words,'' {2017},
  \url{https://www.noswearing.com/dictionary}.

\bibitem{gomes2019machine}
H.~M. Gomes, J.~Read, A.~Bifet, J.~P. Barddal, and J.~Gama, ``{Machine Learning
  for Streaming Data: State of the Art, Challenges, and Opportunities},''
  \emph{ACM SIGKDD Explorations Newsletter}, vol.~21, no.~2, pp. 6--22, 2019.

\bibitem{weka}
``{\em WEKA: The workbench for machine learning},''
  \url{https://www.cs.waikato.ac.nz/ml/weka/}, 2020.

\bibitem{moa-website}
``{\em MOA: Machine Learning for Data Streams},''
  \url{https://moa.cms.waikato.ac.nz/}, 2020.

\bibitem{rajadesingan2015sarcasm}
A.~Rajadesingan, R.~Zafarani, and H.~Liu, ``{Sarcasm Detection on Twitter: A
  Behavioral Modeling Approach},'' in \emph{WSDM}.\hskip 1em plus 0.5em minus
  0.4em\relax ACM, 2015, pp. 97--106.

\end{thebibliography}
